\documentclass[11pt,a4paper]{article}

\usepackage{graphicx}
\usepackage{subcaption}
\usepackage{xcolor}
\usepackage{amsmath} 
\usepackage{amssymb} 
\usepackage{cite}
\usepackage{float}
\usepackage{braket}
\usepackage[T1]{fontenc}
\usepackage{titlesec}
\titleformat{\paragraph}
{\normalfont\normalsize\bfseries}{\theparagraph}{1em}{}
\titlespacing*{\paragraph}
{0pt}{3.25ex plus 1ex minus .2ex}{1.5ex plus .2ex}
\newcommand{\be}{\begin{equation}}
\newcommand{\ee}{\end{equation}}
\newcommand{\bpm}{\begin{pmatrix}}
\newcommand{\epm}{\end{pmatrix}}

\newcommand{\dd}{\mathrm{d}}

\usepackage{tabularx} 
\usepackage{psfrag} 

\usepackage{eurosym} 
\def\?{\euro{}}
\usepackage{footnote}
\usepackage{tablefootnote}

\usepackage{color} 
\usepackage{mathtools}

\definecolor{linkcolor}{rgb}{0,0,0.6} 
\usepackage[pdftex,colorlinks=true,
pdfstartview=FitV,
linkcolor= linkcolor,
citecolor= linkcolor,
urlcolor= linkcolor,
hyperindex=true,
hyperfigures=false]
{hyperref} 

\usepackage{fancyhdr} 

\numberwithin{equation}{section}
\begin{document}

\begin{titlepage}

	\begin{center}

	\vskip .5in 
	\noindent

	{\Large \bf{Non-supersymmetric M-theory vacua and dark bubbles}}

	\bigskip\medskip

	 Vincent Menet,$^1$ Alessandro Tomasiello,$^{1}$ Vincent Van Hemelryck$^{2,*}$ \\

	\bigskip\medskip
	{\small 
$^1$ Dipartimento di Matematica, Universit\`a di Milano-Bicocca, \\ Via Cozzi 55, 20126 Milano, Italy \\ and \\ INFN, sezione di Milano-Bicocca
\\	
	\vspace{.3cm}
$^2$ Department of Physics and Astronomy, Uppsala University, Box 524, SE-75120, Uppsala, Sweden\\
\vspace{.1cm}
$^*$ Present address: Max-Planck-Institut f\"ur Physik (Werner-Heisenberg-Institut),
Boltzmannstraße 8,\\
85748 Garching, Germany
}

	\vskip .5cm 
	{\small \tt  vincent.menet@unimib.it,
    alessandro.tomasiello@unimib.it,
    vincent.vanhemelryck@physics.uu.se/vinvh@mpp.mpg.de}
	\vskip .9cm 
	\end{center}

	\noindent
An expanding bubble in AdS$_5$ has a dS$_4$ geometry; the dark bubble scenario explores whether
this can be our Universe.  Motivated by this, we consider new non-supersymmetric AdS$_5$ compactifications of M-theory and investigate their non-perturbative instabilities. We focus on a compactification on the flag manifold. In certain regions of the flux landscape, the non-supersymmetric vacua become unstable under the nucleation of M5-brane bubbles wrapping an internal two-cycle, thus realising the dark bubble scenario. The cosmological constant on the bubble can
be made parametrically small by tuning the internal fluxes, making the M5-branes close to the extremality bound. The solution relates the induced cosmological constant on the brane and the typical largest length scale of the extra dimensions through $R^{-1}\sim \Lambda_4^{1/4}$ in 4d Planck units, allowing for extra dimensions of the order of tens of microns. We also consider various alternatives, which for various reasons are currently not as successful.
	\noindent

	\vfill
	\eject

	\end{titlepage}

\tableofcontents

\section{Introduction}

The existence of de Sitter vacua among string compactifications is a long-standing debate in string theory. An intriguing observation is that vacuum decay in AdS$_5$ creates an expanding bubble with an induced four-dimensional accelerated cosmology. If the bubble could host four-dimensional Einstein gravity, we would obtain an alternative to compactification for de Sitter. 

This is the so-called \textit{dark bubble} scenario, first put forward in \cite{Banerjee:2018qey}. Many of its physical implications have been studied in \cite{Banerjee:2020wov,Banerjee:2020wix,Banerjee:2021qei,Danielsson:2022fhd,Banerjee:2022ree,Danielsson:2024frw,Basile:2023tvh,Basile:2025lwx,Danielsson:2026qch}. Four-dimensional Einstein gravity (at long distances) is argued to be realised differently from other braneworld models, like RS2 \cite{Randall:1999vf} and KR \cite{Karch:2000ct}, and relies on appropriately chosen mixed boundary conditions at a cut-off brane near the boundary of AdS$_5$ \cite{Banerjee:2023uto,Danielsson:2025ffk,Danielsson:2026wqp}. These are not the standard boundary conditions used for AdS/CFT, and one might perhaps worry whether they are allowed in a theory of quantum gravity. 

Nevertheless, in this paper we are going to take this interesting mechanism at face value, and explore ways to realise it in string theory. Indeed, it is important to establish whether a UV-complete theory of gravity like string theory can even host dark bubbles where the cosmological constant is small in four-dimensional Planck units. In string theory, such bubbles typically arise from brane nucleation, when the brane is super-extremal. The induced cosmological constant is controlled by how far the brane is from being extremal: in particular, the brane must be slightly super-extremal to induce a small cosmological constant. In string theory compactifications, such setups are rather rare, either because super-extremal bubbles do not exist, or because they are too super-extremal. In refs.~\cite{Danielsson:2022lsl,Danielsson:2023alz}, slightly super-extremal D3-branes in a supersymmetry-breaking type IIB background were considered, although the analysis relies on $\alpha'$ corrections that can be difficult to control. Ref.~\cite{Basile:2025lek} considered a construction using slightly super-extremal D3-branes in the non-supersymmetric type 0'B string theory; the authors argued this bubble to be embedded within a non-phenomenologically viable slowly accelerating four-dimensional spacetime.  

In this paper, we consider the dark bubble scenario realised by unstable M5-branes in M-theory compactifications. 
Our main strategy (Sec.~\ref{sec:flag}) is inspired by the (non-supersymmetric) solution AdS$_5\times$ K\"ahler--Einstein$_6$ ($\mathrm{KE}_6$) in eleven-dimensional supergravity \cite{Dolan:1984hv,Pope:1988xj}. A conceivable decay channel would be mediated by an M5 wrapping a two-cycle in $\mathrm{KE}_6$, but this was checked to be absent early on \cite{Martin:2008pf}. Here we observe that one of these $\mathrm{KE}_6$, the flag manifold $F_6 \equiv \mathbb{F}(1,2;3)$, is part of a three-parameter space of homogeneous metrics (including overall size).\footnote{Other irreducible compact six-dimensional coset spaces do not have this feature; $\mathbb{CP}^3$ does have two parameters, but that is not enough to tune the cosmological constant, as we will see.} There are two flux quanta $n_i$ for the four-form $G_4$; by tuning their ratio, we can turn on the M5 channel, and tune the cosmological constant $\Lambda_4$ on the bubble. More precisely, the regime of interest in this model is where $n_1/n_2$ is at a small distance from a certain irrational value. With the help of standard number theory results, we estimate that, as $\Lambda_4$ is made small, the size of the extra dimensions grows as $R^{-1} \sim \Lambda_4^{1/4}$ in 4d Planck units. This relation was postulated in \cite{Montero:2022prj} by combining swampland ideas and compatibility with observations; here it appears naturally. (It also appears in this context in \cite[(8)]{Danielsson:2026wqp}.) Concretely, the observed $\ell_4^2\Lambda_4 \simeq 7.1\times10^{-121}$ leads to extra dimensions of order a few tens of microns. 

We briefly consider a second strategy (Sec.~\ref{sec:7d}) based on AdS$_5$ solutions that are the dimensional reduction of seven-dimensional maximal gauged supergravity on a Riemann surface $\Sigma$, which is a consistent truncation from eleven-dimensional M-theory. We restrict to $\Sigma=S^2$ in order to avoid moduli, which are dangerous in view of possible stringy corrections. However, we find the remaining solutions to be perturbatively unstable.
We nevertheless briefly consider the behaviour of a would-be M5-brane wrapping the $S^2$. This time we achieve a less favourable hierarchy of scales than in our flag manifold model: $R^{-1} \sim \Lambda_4^{1/2}$, in four-dimensional Planck units. 
We obtain the same hierarchy of scales for a final dark bubble construction in Appendix \ref{sec:kahler}, embedded in  AdS$_5\times\text{K\"ahler}_4\times \Sigma$ and AdS$_5\times \Sigma\times\Sigma\times \Sigma$ solutions of eleven-dimensional supergravity.

\section{The dark bubble scenario}
\label{sec:rev}
Given the  challenge of establishing trustworthy meta-stable de Sitter solutions from string compactifications, the dark bubble scenario \cite{Banerjee:2018qey} aims to provide a braneworld alternative by considering an accelerated FLRW cosmology on an expanding brane in a spacetime of one dimension higher.

\subsection{The five-dimensional perspective}

In its simplest incarnation, the dark bubble arises as a non-perturbative decay of some AdS$_5$ vacuum with inverse AdS radius $k_+\equiv 1/L_+$ to a more stable AdS$_5$ vacuum with inverse radius $k_-\equiv 1/L_-$, $\Lambda_5=-6/L^2$. The decay is mediated by a three-brane, with tension $\sigma$. When looking at such spherical bubbles in global AdS coordinates, the Israel junction conditions dictate how the two AdS$_5$ spacetimes are glued together across the brane, and they must satisfy
\begin{equation}
	\sigma= \frac{3}{\ell_5^3 a} \left(\sqrt{k_-^2 a^2+1 + \dot{a}^2} - \sqrt{k_+^2 a^2+1 + \dot{a}^2} \right) 
\end{equation}
where $a$ is the radius of the bubble and also the scale factor of the FLRW induced metric on the brane.
The 5d Planck scale is taken to be the same across the bubble.
Then, assuming that the radius of the bubble is much larger than the AdS radii on either side, i.e.~$k_\pm a \gg 1$, the junction condition can be expanded and arranged such that it takes the form of the Friedmann equation:
\begin{equation}
	H^2 = \frac{\dot{a}^2}{a^2} = \frac{\ell_4^2}{3}\rho_\Lambda - \frac{1}{a^2} + \cdots
\end{equation}
The 4d Planck scale $\ell_4$ and the dark energy density $\rho_\Lambda$ are given by
\begin{equation}\label{eq:l45}
	\ell_4^2 = 2 \ell_5^3 (k_+^{-1}-k_{-}^{-1})^{-1} = 2 \ell_5^3 \frac{k_- k_+}{k_- - k_+}\,, \qquad \rho_\Lambda = \sigma_\mathrm{crit} - \sigma\,,
\end{equation}
and the critical tension is given by the difference of inverse AdS radii: 
\begin{equation}\label{eq:sigma-crit}
	\sigma_\mathrm{crit} = \frac{3}{\ell_5^3}(k_- - k_+)\,.
\end{equation}
From now on, we write tension in terms of the critical tension as $\sigma = \sigma_\mathrm{crit}(1 - \epsilon)$. Using this, 
\begin{equation}\label{eq:lambdaeps}
	\rho_\Lambda = \sigma_\mathrm{crit} \; \epsilon = 12  \frac{k_-^2 k_+^2}{k_- - k_+} \; \frac{\ell_5^3}{\ell_4^{4}}\;\epsilon\,.
\end{equation}
Note that we have rewritten this to express this quantity in the relevant 4d Planck units (and the inverse AdS radii in 5d Planck units). The dark energy density is therefore positive when $\epsilon > 0$ and is small in 4d Planck units whenever
\begin{equation}
	\ell_5^3\frac{k_-^2 k_+^2}{k_- - k_+}\; \epsilon \ll 1\,.\label{eq:ded}
\end{equation}
If one wants this to touch upon the dark energy density of our universe, $\ell_4^4\rho_\Lambda$ has to approach $10^{-120}$.

A crucial question about this model is whether gravity can be localised to the bubble, yielding an effectively 4d gravitational potential. A first idea was to use strings ending on it \cite{Banerjee:2019fzz,Banerjee:2020wix,Banerjee:2020wov}.  A more recent strategy uses a cut-off brane near the AdS boundary, with appropriately chosen mixed boundary conditions~\cite{Banerjee:2023uto,Danielsson:2025ffk}.

One also wants to couple this model to gauge fields. This has been the subject of a series of subsequent works, see refs.~\cite{Basile:2023tvh, Danielsson:2024frw,Basile:2025lek,Danielsson:2026qch} for the inclusion of electromagnetism (see also \cite{Basile:2025lwx} for non-Abelian gauge fields).
For short introductions to the topic, see \cite{Banerjee:2022ree} and the more recent \cite{Danielsson:2026wqp}. 

An important question is whether there is a UV theory of gravity in which the dark bubble model can be embedded. In the context of string theory, this question reduces to the existence of a compactification giving rise to unstable AdS$_5$ vacua that decay non-perturbatively through brane nucleation. Such vacua must be non-supersymmetric; hence this question is related to a well-known conjecture \cite{Ooguri:2016pdq} according to which all non-supersymmetric AdS solutions are unstable.
In such string compactifications, the nucleated branes are typically D-branes, which are charged under RR fields, and induce a shift in the RR flux when crossing them, giving rise (through moduli stabilisation) to shifts in the inverse AdS radius $k$. 

In the seminal paper \cite{Banerjee:2018qey}, the authors investigated a dark bubble realisation in the context of the non-supersymmetric Romans vacuum \cite{Romans:1984an} of type IIB string theory, where supersymmetry is broken by non-trivial three-form flux. There, the decay is mediated by a $(p,q)$-five-brane. However, that particular compactification does not lead to a small dark energy density.
Embeddings in the non-supersymmetric type 0'B string theory also have other phenomenological issues \cite{Basile:2025lek}. 
So far, the embedding of a supersymmetry-breaking $\mathrm{AdS}_5 \times S^5$ black hole in type IIB string theory \cite{Danielsson:2022lsl, Danielsson:2023alz} looks the most promising, although the sub-extremality parameter $\epsilon$ is argued to arise from $\alpha'$ corrections, of which a full picture is still lacking.

In what follows, we aim to explore the dark bubble scenario in the context of M-theory compactifications.

\subsection{An (anti-)M5-brane realisation}

In the context of M-theory, a natural realisation of the dark bubble scenario would be to consider non-supersymmetric vacua of the form AdS$_5\times M_6$, for which the simplest incarnation of the four-dimensional bubble would be an (anti-)M5-brane, being a domain wall in AdS$_5$ while wrapping a two-cycle $\Sigma$ inside $M_6$. 

We shall now derive the criterion for a given AdS$_5\times M_6$ solution of M-theory to undergo the decay process mediated by a probe M5-brane, and then identify the five-dimensional tension of this brane. 

The condition for such an M5-brane to nucleate and expand can be thought of as the condition for the existence of a Euclidean instanton\cite{Coleman:1980aw,Brown:1988kg}. Here we restrict our study to domain walls that are spherically symmetric in AdS$_5$, and for which we take the Euclidean metric
\begin{equation}\label{eq:eads}
    \dd s^2_{\rm EAdS_5}= L^2 (\dd r^2 +\sinh^2 r \dd s^2_{S^{4}})\,.
\end{equation}
The (Euclidean) M5-brane action is, as usual,
\be \label{eq:M5_Action}
S=T_\text{M5}\left(\int \dd^{6} \sigma \sqrt{g}- \int C_6\right),\ee
with $T_\text{M5}$ the brane tension and for which we make the gauge choice
\be 
C_6=L^5 c(r) \text{Vol}(S^4)\wedge \ast_6 G\, , 
\ee
with $c'(r)=\sinh^4 r$. Next, we introduce an effective five-dimensional field strength $F_5\equiv E\text{Vol}(\text{AdS}_5)=\frac{h}{L}\text{Vol}(\text{AdS}_5)$, with $h$ the magnitude of this top-degree field. The effective tension and charge density of the brane are then
\begin{equation}\label{eq:tq}
    \sigma = T_{\rm M5} \text{Vol}(\Sigma) \,, \qquad
q= T_{\rm M5}\frac L h \int_\Sigma\ast_6 G \,.
\end{equation}
Plugging this into the action \eqref{eq:M5_Action} results in
\begin{equation}\label{eq:M5_Euclidean_Action}
    S=T_\text{M5}L^{4} \text{Vol}(S^4)\left(\sinh^{4} r\, \text{Vol}(\Sigma)-L\, c(r)\int_\Sigma\ast_6 G\right)\,
\end{equation}
with $\Sigma$ the internal two-cycle wrapped by the M5-brane. Extremising \eqref{eq:M5_Euclidean_Action} in $r$, we find that there exists an instanton solution if there is a $r_0$ such that
\be 
\tanh{r_0}=\frac{4\text{Vol}(\Sigma)}{L\int_\Sigma\ast_6 G}\, ,\label{eq:r0}
\ee
and $r_0$ is then the radius of the instanton. This is only realised when we have 
\be
\rho\equiv\frac{qh}{4 \sigma} = \frac{L\int_\Sigma\ast_6 G}{4\text{Vol}(\Sigma)}>1\, .\label{eq:rho}
\ee

We will refer to this quantity $\rho$ as the \emph{stability ratio}. Roughly speaking the condition \eqref{eq:rho} ensures that the repulsive force on the brane, related to its charge, wins over its contracting gravitational force.

It is useful to connect the effective  charge density $q$ of \eqref{eq:tq} with the critical tension of the brane \eqref{eq:sigma-crit}. This can be done in the spirit of \cite{Brown:1988kg} as follows. Separating the contribution from the five-dimensional field strength $F_5$ from everything else, we write
\begin{equation}
    \Lambda_5= \Lambda^{(0)}_5 +\frac14 E^2\,.\label{eq:BT}
\end{equation}
The flux equations of motion give $\Delta E = 2\ell_5^3 q$ across the brane. In the probe approximation, this variation is much smaller than the existing flux, so we can keep the term linear in $q$ and use \eqref{eq:BT} to write $\Delta \Lambda_5= \ell_5^3 q E$. Recalling $E=\frac{h}{L}$ and $\Lambda_5=-6 k^2 = -6/L^2$, we find\footnote{Here we define $\Delta k=k_--k_+$.}
\be \label{eq:repro-sigma}
\frac{qh}{4}=\frac{3\Delta k}{\ell_5^3}\equiv\sigma_\text{crit}, 
\ee
reproducing \eqref{eq:sigma-crit}. The quantities on the left-hand side should be understood as evaluated \emph{outside} the bubble. From \eqref{eq:rho}, we can then connect the stability ratio with the deviation to criticality of the brane $\epsilon$ as follows:
\be
\rho=\frac{\sigma_\text{crit}}{\sigma}=\frac{1}{1-\epsilon}.\label{eq:eps}
\ee
In the same vein, one can also reproduce the calculation of $\Lambda_4$ from the five-dimensional perspective of eq.~\eqref{eq:lambdaeps}. Equations~\eqref{eq:r0}-\eqref{eq:rho} give us $\tanh r_0 = 1/\rho$; from eq.~\eqref{eq:eads} we see $L_{{\rm dS}_4}=L \sinh r_0= L(\rho^2-1)^{-1/2}\simeq L/\sqrt{2\epsilon}$. So
\begin{equation}\label{eq:Lambda4-probe}
    \frac{\rho_\Lambda}{m_4^4}= \frac{\Lambda_4}{m_4^2}= 3\frac{\ell_4^2}{L_{{\rm dS}_4}^2}= 6 \frac{\ell_4^2}{L^2}\epsilon\,,
\end{equation}  
which one can also obtain by combining eq.~\eqref{eq:l45} and \eqref{eq:lambdaeps} in the limit where $k_+\simeq k_- \equiv 1/L$. 

The five-dimensional and probe perspectives have complementary strengths. The former has the advantage that it can apply to decays connecting  arbitrarily different vacua; the latter can resolve the details of the internal dimensions, which the former cannot. Ultimately the only purely five-dimensional ingredient one needs to characterise the dark bubble is the way $\ell_4$ and $\ell_5$ relate via \eqref{eq:l45}. 

In practice, we will evaluate the charge $qh$ and tension $\sigma$ for some M5-branes in eleven-dimensions, and extract the deviation to criticality $\epsilon$ of the corresponding four-dimensional bubble via \eqref{eq:repro-sigma} and \eqref{eq:eps}. We emphasise that the dark bubble analysis uses a fixed five-dimensional Planck scale across the brane, so the M5-brane charge and tension must be expressed in the corresponding five-dimensional Planck units.

Coming back to the dark bubble scenario, a small four-dimensional dark energy density requires a small, positive $\epsilon$, via \eqref{eq:ded}.  Being able to tune the smallness of $\epsilon$ represents one of the main challenges of higher dimensional dark bubble constructions. In what follows, we derive new non-supersymmetric solutions of eleven-dimensional supergravity that admit M5-branes with arbitrarily small deviation to criticality $\epsilon$.

\section{The flag model} 
\label{sec:flag}

In this section we will consider quantum instabilities of the AdS$_5\times F_6$ spacetime, where 
\begin{equation}\label{eq:su3/u1u1}
		F_6 = \mathbb{F}(1,2;3)\equiv \mathrm{SU}(3)/\mathrm{U}(1)\times \mathrm{U}(1)
\end{equation}
is called a flag manifold, for reasons to be reviewed below. It has appeared in IIA flux vacua in \cite{Tomasiello:2007eq,Koerber:2008rx,Menet:2025nbf}. 

We focus on this space because it has three free parameters $\alpha_i$, the largest number among all six-dimensional (compact irreducible) homogeneous spaces. As we will review shortly, the Betti number $b_2(F_6)=2$, and this will let us introduce two flux parameters $f_a$. The Einstein equations will fix three combinations of these, leaving two free variables. Flux quantisation will discretise these two in terms of two rational combinations of three flux quanta. One of these rational combinations is what will allow us to tune the stability ratio (and thus the four-dimensional cosmological constant). 

The AdS$_5\times \mathrm{KE}_6$ solutions \cite{Dolan:1984hv,Pope:1988xj} can have many more moduli, but M5-branes are stable for them \cite{Martin:2008pf}. This finding does not apply to homogeneous spaces, but in that case tuning the stability ratio is impossible with fewer than the three parameters that $F_6$ has. Consider for example $\mathbb{CP}^3$: it has a \emph{two}-parameter homogeneous metric, and \emph{one} internal two-cycle. One obtains only two isolated solutions: one with an internal metric KE$_6$, the other with a squashed twistor fibre \cite{Imaanpur:2012jx}. We will recover them below in passing, exploiting the close formal relation between $\mathbb{CP}^3$ and $F_6$; we will conclude that M5 instabilities are absent for both. 

We will begin with some geometry preliminaries in Sec.~\ref{sub:geo}. We will then present the solution in Sec.~\ref{sub:sol}. In Sec.~\ref{sub:m5} we will analyze its M5-stability, and the consequences for the dark bubble scenario. 

\subsection{Flag geometry} 
\label{sub:geo}

$F_6$ has a remarkably rich geometry, and can be viewed through many lenses. It is a homogeneous space, but it has other descriptions that are useful in some subloci of its parameter space: as the total space of the twistor bundle of $\mathbb{CP}^2$, as a nearly-K\"ahler manifold, as a K\"ahler manifold, as a locus of degree $(1,1)$ in $\mathbb{CP}^2\times \mathbb{CP}^2$. 

Most of the aspects we need can be extracted from its homogeneous structure \eqref{eq:su3/u1u1}. We need some useful techniques that apply to all $G/H$ coset spaces; we will review them along the way, beginning with the theory of invariant forms. For more details, see e.g.~\cite[Sec.~4.4.2]{Tomasiello:2022dwe} or \cite[Sec.~3.2]{Raucci:2025bev}.

\subparagraph{Forms.} Let the Lie algebra $\mathfrak{g}$ of $G$ have generators $T_I$, with commutators $[T_I,T_J]= f^K{}_{IJ} T_K$. Among these, we call $T_\alpha$ the generators of the Lie subalgebra $\mathfrak{h}$, of the subgroup $H$, and we choose the remaining $T_i$ such that $f^\alpha{}_{i \beta}=0$; this can always be done when $H$ is compact. 

The one-forms $g^{-1}\dd g = \lambda^I T_I(= \lambda^i T_i + \lambda^\alpha T_\alpha)$ are forms on $G$; they are invariant under left multiplication by elements of $G$, and satisfy the Maurer--Cartan equations $\dd \lambda^I = - \frac12 f^I{}_{JK} \lambda^J \wedge \lambda^K$. For a form $\omega$ on $G$ to descend to a form on $G/H$, it should be horizontal under the projection map $\pi: G \to G/H$, and invariant under it: $\iota_v \omega= L_v \omega=0$, for $v$ vertical. The first property is satisfied by the $\lambda^i$ and their wedge products $\lambda^{i_1\ldots i_k}\equiv \lambda^{i_1} \wedge \ldots \wedge \lambda^{i_k}$. The second property is then $L_v \omega = \{ \dd, \iota_v\} \omega = \iota_v \dd \omega=0$, which means $\dd \omega$ should also be horizontal. In practice this means that we look for a linear combination of the $\lambda^{i_1\ldots i_k}$ whose exterior derivative does not contain any $\lambda^\alpha$.

We now get to our case $G=\mathrm{SU}(3)$, $H=\mathrm{U}(1)\times \mathrm{U}(1)$. We take ${f^1}_{54}={f^1}_{36}={f^2}_{46}={f^2}_{35}={f^3}_{47}={f^5}_{76}=1$, ${f^1}_{27}=2$, ${f^3}_{48}={f^5}_{68}=\sqrt{3}$ and the further identities obtained by cyclically permuting the indices; this is just a multiple of the usual Gell-Mann choice. The independent invariant forms are 
\begin{subequations}
\begin{align}
	&j_1= \lambda^{12}\,,\qquad j_2=- \lambda^{34}\,,\qquad j_3= \lambda^{56}\,,\\
	&\psi=\lambda^{135}+\lambda^{146}-\lambda^{236}+\lambda^{245}\,, \qquad \Tilde{\psi}=-\lambda^{136}+\lambda^{145}-\lambda^{235}-\lambda^{246}\,,
\end{align}
\end{subequations}
and their wedge products. They satisfy
\begin{subequations}
\begin{align}
	&j_1^2=j_2^2=j_3^2=0 \, ,\qquad j_1 \wedge j_2 \wedge j_3 =-\frac14 \psi \wedge \tilde \psi\,;\\
	\label{eq:d-forms}
	&\dd j_i=\psi \, ,\qquad \dd\tilde{\psi}=4(j_1\wedge j_2+j_1\wedge j_3+j_2\wedge j_3)\,.
\end{align}
\end{subequations}

\subparagraph{Hypersurface, twistor bundle.} 

Consider $i$-dimensional vector subspaces $l_i\subset \mathbb{C}^3$. A (complete) flag in $\mathbb{C}^3$ is a pair $(l_1,l_2)$ such that $l_1 \subset l_2$. Now $F_6$ can be written as the space of such flags: 
\begin{equation}\label{eq:F6-flag}
	F_6= \{ (l_1,l_2)\ | \  l_1 \subset l_2\subset \mathbb{C}^3 \}\,.
\end{equation}
Indeed, SU(3) acts on this space of flags, and the stabiliser of any flag is $\mathrm{U}(1)\times \mathrm{U}(1)$. 

The space of $l_1= \mathrm{Span}(v)$'s is famously $\mathbb{CP}^2$. Writing an $l_2$ as $\{ w_i z_i=0\}$ for some parameters $w_i$, we see that the space of $l_2$'s is also a $\mathbb{CP}^2$. Demanding $l_1\subset l_2$ gives 
\begin{equation}\label{eq:F6-hyp}
	F_6=\{v_i w_i = 0\}\subset \mathbb{CP}^2\times \mathbb{CP}^2\,.
\end{equation}
There are now two obvious projections to $\mathbb{CP}^2$, but there is also a third \emph{twistor fibration} defined by $\pi:(l_1,l_2)= \epsilon_{ijk}v_j^* w_k$, where the right-hand side is viewed as the homogeneous coordinates in a third $\mathbb{CP}^2$ (see for ex.~\cite[Sec.~5]{Altavilla_2022}). The fibre is an $S^2$: we can write
\begin{equation}\label{eq:F6-tw}
	S^2\hookrightarrow F_6 \to \mathbb{CP}^2\,.
\end{equation}
The triality between the three above maps to $\mathbb{CP}^2$ is related to the Weyl symmetry of SU(3).

\subparagraph{Metric.} Just like for forms, a metric on $G/H$ can again be defined as a metric on $G$ that descends to $G/H$. The invariance condition now reads $g_{kj}f^k{}_{i \alpha}+g_{ik}f^k{}_{j \alpha}=0$. For our case this gives
\begin{equation}\label{eq:ds2-F6}
	\dd s^2_{F_6} = \alpha_1^2 (\lambda^1\lambda^1+\lambda^2\lambda^2)+\alpha_2^2 (\lambda^3\lambda^3+\lambda^4\lambda^4)+\alpha_3^2 (\lambda^5\lambda^5+\lambda^6\lambda^6)\,.
\end{equation}
This could also be obtained by building an SU(3)-structure $(J,\Omega)$ from the $j_i$ and $\psi$. 

We will sometimes find it useful to separate the $\alpha_a$ into a volume parameter $\alpha_3$ and two shape parameters $s_{1,2} \equiv \alpha_{1,2}^2/\alpha_3^2$; in other words,
\begin{equation}\label{eq:par}
	\alpha_1^2 = \frac{R^2}{s_1} \, ,\qquad\alpha_2^2 = \frac{R^2}{s_2} \, ,\qquad \alpha_3^2 = R^2\,.
\end{equation}

While they will not be too important for us, let us also mention various special loci in the space of the $\alpha_a$:
\begin{itemize}
	\item When $\alpha_2^2= \alpha_1^2 + \alpha_3^2$, the two-form $J=- \alpha_1^2 j_1 + \alpha_2^2 j_2 + \alpha_3^2 j_3$ is closed, and defines a K\"ahler structure; as will become clear later, it is the pull-back of the one on $\mathbb{CP}^2\times \mathbb{CP}^2$ under the inclusion in \eqref{eq:F6-hyp}. There are similar K\"ahler structures for $\alpha_1^2= \alpha_2^2 + \alpha_3^2$ or for $\alpha_3^2= \alpha_1^2 + \alpha_2^2$. 
	\item When the $\alpha_a$ are all equal, one can introduce a \emph{nearly K\"ahler} structure: an SU(3)-structure $(J,\Omega)$ such that $\dd J = 3 \mathrm{Im} \Omega$, $\dd \mathrm{Re} \Omega = 2 J^2$. 
	\item When two of the $\alpha_a$ are equal, the metric is the one induced by the twistor fibration \eqref{eq:F6-tw}, with the Fubini--Study metric on $\mathbb{CP}^2$. In this case, the properties of the space are essentially identical to those of $\mathbb{CP}^3$, the twistor fibration over $S^4$, even though topology does change, as $h_2(\mathbb{CP}^3)=1$. If we take $\alpha_1=\alpha_2$, \eqref{eq:par} gives $s_1=s_2\equiv s$, which is the parameter called $\sigma$ in \cite{Tomasiello:2007eq} and \cite[Sec.~4.1]{Menet:2025nbf}.
\end{itemize}

\subparagraph{(Co)homology.}
From \eqref{eq:d-forms} we see that both $j_1-j_2$, $j_2-j_3$ are closed two-forms, that are also co-closed relative to the metric \eqref{eq:ds2-F6}. We do not find any odd harmonic forms. Since the homogeneous sspace is compact, all harmonic forms are invariant.

To compute the cohomology, we see from the point of view of \eqref{eq:F6-hyp} that there are two independent four-cycles $H$, $\tilde H$, defined as the intersection of $F_6$ with a (linear) hyperplane in the first and second $\mathbb{CP}^2$ respectively. 
 So $h_4=h_2=2$.  Counting intersections we see that $H^3=\tilde H^3=0$, $H^2 \cdot \tilde H= H \cdot\tilde H^2=1$. Familiar algebraic geometry techniques (Lefschetz hyperplane and Atiyah--Singer index theorems) then give $h_1=h_3=0$. The same results can also be obtained from \eqref{eq:F6-tw} using the Gysin exact sequence: from this viewpoint, $H_2(F_6)$ is generated by the single two-cycle of the base and by the fibre. 

For our purposes, we can also introduce explicit two-cycles in the following way. Consider for example the subalgebra $\{T_1,\,T_2,\,T_7\}$, which is the $\mathfrak{su}(2)$ associated to one of the simple roots $r_1$ of SU(3). Exponentiating it gives an $\mathrm{SU}(2)\subset \mathrm{SU}(3)$; $T_7\in \mathfrak{u}(1)\oplus \mathfrak{u}(1)$, so we obtain an $\mathrm{SU}(2)/\mathrm{U}(1)\cong S^2\subset F_6$, which we call $S_1$. There are two additional $\mathfrak{su}(2)$ subalgebras, associated to roots $-r_1-r_2$, $r_2$ (with generators $\{T_4,\,T_3,\,T_7+ \sqrt{3}T_8\}$ and $\{T_5,\,T_6,\,-T_7+ \sqrt{3}T_8\}$). They exponentiate to two additional two-spheres $S_2$, $S_3$. From the viewpoint \eqref{eq:F6-flag}, the action defining $S_1$ ($S_3$) only moves the $l_1$ ($l_2$) subspace; it is then a linear curve in the first (second) $\mathbb{CP}^2$ in \eqref{eq:F6-hyp}. In other words:\footnote{A different way to work out the cohomology and intersections is via the more general theory of Schubert cycles, which are associated combinatorially to permutations. A related result is the Borel presentation theorem, which states in this case that the cohomology ring is isomorphic to $\mathbb{Z}[x_1,x_2,x_3]/I$, where $I$ is the ideal generated by symmetric functions up to degree three: $x_1+x_2+x_3$, $x_1 x_2 + x_2 x_3 + x_3 x_1$, $x_1 x_2 x_3$. For a recent review, see \cite{billey-gao-pawlowski}.}
\begin{equation}\label{eq:SH}
	S_1 = H \cdot H \, ,\qquad S_3 = \tilde H \cdot \tilde H\,.
\end{equation}

For $g_1$ a general element of the first SU(2), we see $\mathrm{Tr}(g_1^{-1}\dd g_1 T_{3,4,5,6})=0$; so the pull-back of $j_2$, $j_3$ to $S_1$ is zero. $j_1$ gives instead its volume form, with normalisation $\int_{S_1} j_1 = \pi$. Similar considerations lead to
\begin{equation}\label{eq:intSj}
	\int_{S_a} j_b = \pi \delta_{ab}\,.
\end{equation}
The $j_a$ are not closed, but we can define $h_1\equiv \frac1\pi (j_2-j_3)$, $h_2\equiv \frac1\pi (j_3-j_1)$, $h_3\equiv \frac1\pi (j_1-j_2)$, which are closed and normalised so that any two of them are generators of $H^2(F_6,\mathbb{Z})$. 

Recall now that Poincar\'e duals are defined by $\int_S \alpha = \int_{F_6} \alpha \wedge \mathrm{PD}(S)$ $\forall \alpha$ in cohomology, $\int_S \alpha = S \cdot \mathrm{PD}(\alpha)$ $\forall S$ in homology. Using these repeatedly on the bases $\{S_1,S_3\}$ of $H_2$ and $\{h_1,h_2\}$ of $H^2$, we find\footnote{For example, $0=H \cdot S_1 = \int_{S_1} h_1$, $1=H \cdot S_3 = - \int_{S_3} h_1$ leads to the first in \eqref{eq:PDH}. \eqref{eq:SH} gives the first in \eqref{eq:PDS}, which can also be confirmed via $1=\int_{S_1} h_3 = \int h_3 \wedge \mathrm{PD}(S_1)=\frac2{\pi^3}\int \lambda^{1\ldots 6}$, which holds because $\mathrm{Vol}(F_6)=\frac{\pi^3}2$ for the $\alpha_a=1$ metric \cite[(7.14)]{Boya:2003km}.}
\begin{subequations}
\begin{align}
	\label{eq:PDH}
	&\mathrm{PD}(H)= - h_1 \, ,& & \mathrm{PD}(\tilde H) = h_3 \, ;\\
	\label{eq:PDS}
	&\mathrm{PD}(S_1) = -\frac2{\pi^2} j_2 \wedge j_3 \, ,& & 
	\mathrm{PD}(S_2) = -\frac2{\pi^2} j_1 \wedge j_3 \, ,\qquad
	\mathrm{PD}(S_3) = -\frac2{\pi^2} j_1 \wedge j_2 \, .
\end{align}
\end{subequations}
The second of \eqref{eq:d-forms} now implies
\begin{equation}
	\sum_a S_a = 0 
\end{equation}
in homology. 

One also expects an analogue of \eqref{eq:intSj} for four-cycles. One expects four-\textit{chains} $D_i$ such that $\int_{D_i} j_j \wedge j_k \neq 0$ iff $i$, $j$, $k$ are all different, such that $\partial D_1=\partial D_2=\partial D_3$ (just like $\dd j_1=\dd j_2 = \dd j_3$). The differences $D_{13}=D_1-D_3$, $D_{23}=D_2-D_3$ would then be four-cycles, $\partial D_{13}=\partial D_{23}=0$. Instead of constructing these $D_i$, we define more directly $D_{13}\equiv -H + \tilde H$, $D_{23}\equiv - H$. From \eqref{eq:PDS} it then follows that 
\begin{equation}\label{eq:int-D13-j}
    -\frac2{\pi^2}\int_{D_{13}}(j_2 \wedge j_3, j_1 \wedge j_3 , j_1 \wedge j_2) = (1,0,-1) \,\qquad 
    -\frac2{\pi^2}\int_{D_{23}}(j_2 \wedge j_3, j_1 \wedge j_3 , j_1 \wedge j_2) = (0,1,-1)\,.
\end{equation}
So indeed everything works according to the rule $\int_{D_i} j_j \wedge j_k = -\pi^2/2$ if $i$, $j$, $k$ are all different, $0$ otherwise.


\subsection{M-theory solution} 
\label{sub:sol}

We can now introduce our Ansatz. The metric is
\begin{equation}
	\dd s^2_{11} = L^2\,\dd s^2_{\mathrm{AdS}_5} + \dd s^2_{F_6}\,,
\end{equation}
where the metric of the flag manifold is given in eq.~\eqref{eq:ds2-F6}.
As for the four-form, spacetime symmetries demand it to be purely internal. We consider compactifications without M-brane sources, such that the equations of motion are $\dd G=0$, $\dd \ast G + \frac{1}{2}G \wedge G=0$, which reduce to the condition that $G$ should be harmonic. As a consequence, $G$ takes the form\footnote{The factor of $R^2$ is natural for dimensional reasons. Recall that $R$ is related to the metric parameters $\alpha_a$ by \eqref{eq:par}. A mild inconvenience of this choice is that it privileges $\alpha_3$, breaking the symmetry among them; we will see however that in fact this symmetry survives by making the $f_i$ transform appropriately.}
\begin{equation}\label{eq:G}
	G = R^2 \ast_6\left(f_1(j_1-j_3)+f_2(j_2-j_3)\right)\,,
\end{equation}
where $f_1$ and $f_2$ can be thought of as flux parameters, but do not represent the quantised fluxes directly. We relate them to the quantised fluxes later in this section.

\subparagraph{Einstein equations.} 

The Einstein equations read $R_{MN}=\frac12 \iota_M G \cdot \iota_N G -\frac16 g_{MN}|G|^2$.\footnote{Given a $k$-form $\alpha_k=\frac1{k!}\alpha_{M_1\ldots M_k} \dd x^{M_1}\wedge \ldots \wedge \dd x^{M_k}$, we define $\iota_M \alpha_k \equiv \frac1{(k-1)!}\alpha_{M M_2\ldots M_k} \dd x^{M_2}\wedge \ldots \wedge \dd x^{M_k} $,  $\alpha_k \cdot \beta_k\equiv \frac1{k!} \alpha_{M_1\ldots M_k} \beta^{M_1\ldots M_k}$. This dot is not to be confused with the intersection of homology classes in Sec.~\ref{sub:geo}.} The Einstein tensor can be found using \cite[(3.36)]{Raucci:2025bev}. The stress-energy tensor of $G$ is easily computed using $j_a \cdot j_b = \delta_{ab}$. 
All in all, the Einstein equations read 
\begin{subequations}\label{eq:eoms}
\begin{align}
\label{eq:eom-L}
	-4 &= -\frac{1}{6}\,L^2 R^4\!\left(\frac{(f_1+f_2)^2}{\alpha_3^4}
	+ \frac{f_2^2}{\alpha_2^4} + \frac{f_1^2}{\alpha_1^4}\right)\,;\\
	\label{eq:eom-R11}
    R_{11}=R_{22}=\frac{\alpha_1^4-\alpha_2^4-\alpha_3^4+6\alpha_2^2\alpha_3^2}{\alpha_1^2\alpha_2^2\alpha_3^2}
	&= \frac{R^4}{6}\!\left(2\frac{(f_1+f_2)^2}{\alpha_3^4}
	+ 2\frac{f_2^2}{\alpha_2^4} - \frac{f_1^2}{\alpha_1^4}\right)\,,\\
    \label{eq:eom-R33}
	R_{33}=R_{44}=\frac{\alpha_2^4-\alpha_1^4-\alpha_3^4+6\alpha_1^2\alpha_3^2}{\alpha_1^2\alpha_2^2\alpha_3^2}
	&= \frac{R^4}{6}\!\left(2\frac{(f_1+f_2)^2}{\alpha_3^4}
	- \frac{f_2^2}{\alpha_2^4} + 2\frac{f_1^2}{\alpha_1^4}\right)\,,\\
	R_{55}=R_{66}=\frac{\alpha_3^4-\alpha_1^4-\alpha_2^4+6\alpha_1^2\alpha_2^2}{\alpha_1^2\alpha_2^2\alpha_3^2}
	&= \frac{R^4}{6}\!\left(-\frac{(f_1+f_2)^2}{\alpha_3^4}
	+ 2\frac{f_2^2}{\alpha_2^4} + 2\frac{f_1^2}{\alpha_1^4}\right)\,.
\end{align}	
\end{subequations}
The first equation represents the external Einstein equation and determines the AdS scale $L$. The last three equations represent the internal Einstein equations, and the Ricci tensor components $R_{11}$, $R_{22}$, etc. are expressed in terms of the coframe basis parametrised by the $\lambda^i$. 
The system is invariant under $f_i\to -f_i$, so solutions always come in pairs. Solving these equations directly 
is rather tedious, so we take a different route. 
We define 
\begin{equation}
    q_1=f_1^2\,, \quad q_2=f_2^2\,, \quad q_3= f_1 f_2\,;
\end{equation}
and substituting this into the last three equations of \eqref{eq:eoms} makes them formally linear in the $q_i$'s. This results in the following solution: 
\begin{subequations}\label{eq:fi-sol}
\begin{align}
	q_1 = f_1^2=&\frac{2 \alpha_1^2}{3 \alpha_2^2 \alpha_3^2 R^4}(\alpha_2^4 -5 \alpha_1^4 + \alpha_3^4 - 6 \alpha_2^2 \alpha_3^2 +12 \alpha_1^2(\alpha_2^2 +	\alpha_3^2)) \,,\\
    q_2 = f_2^2=&\frac{2 \alpha_2^2}{3 \alpha_1^2 \alpha_3^2 R^4}(\alpha_1^4 -5 \alpha_2^4 + \alpha_3^4 - 6 \alpha_1^2 \alpha_3^2 +12 \alpha_2^2(\alpha_1^2 +	\alpha_3^2))\,,\\
    q_3 = f_1 f_2 =& 	\frac1{3 \alpha_1^2 \alpha_2^2 \alpha_3^2 R^4} \Big(5 (\alpha_1^8 +\alpha_2^8) - (12 \alpha_1^4 + 12 \alpha_2^4 + 2 \alpha_1^2 \alpha_2^2) \alpha_1^2 \alpha_2^2\\
     &\hspace{1cm}
      -   6 (2 \alpha_1^6 + 2 \alpha_2^6 - (\alpha_1^2 + \alpha_2^2) \alpha_1^2 \alpha_2^2) \alpha_3^2
     - 6 \alpha_1^2 \alpha_2^2 \alpha_3^4 + 
  12 (\alpha_1^2 + \alpha_2^2) \alpha_3^6 - 5 \alpha_3^8 \Big) \nonumber
.
\end{align}	
\end{subequations}
However, these solutions are subject to the constraint\footnote{When $q_{1,2}$ are positive, there are in principle four possibilities, $f_1=\pm \sqrt{q_1}$, $f_2=\pm \sqrt{q_2}$; the sign of $q_3$ selects two of them, exchanged by the aforementioned $f_i\to -f_i$ symmetry.}
\begin{equation}
    q_3^2=q_1 q_2\,,
\end{equation}
which gives a constraint on the $\alpha_a$. This boils down to
\begin{equation}\label{eq:Sigma}
	\begin{split}
	\Sigma \equiv& -\frac{1}{9}(e_1^2 - 4e_2)(5e_1^3 - 22e_1 e_2)^2 
		- \frac{4}{9}e_1(19e_1^2 - 80e_2)(5e_1^2 - 22e_2)e_3 \\
		&+ 12(-15e_1^2 + 64e_2)e_3^2 = 0 \,,
	\end{split}
\end{equation}
where $e_1 \equiv \alpha_1^2 + \alpha_2^2 + \alpha_3^2$, $e_2 \equiv \alpha_1^2 \alpha_2^2 + \alpha_1^2 \alpha_3^2 + \alpha_2^2 \alpha_3^2$, $e_3\equiv \alpha_1^2 \alpha_2^2 \alpha_3^2$ are polynomials symmetric in the $\alpha_a^2$. In the parametrisation eq.~\eqref{eq:par}, this is only a constraint on the two parameters $s_1$ and $s_2$: all $R$-dependence drops out due to homogeneity. As a summary, at this stage we have solved \eqref{eq:eom-R11}--\eqref{eq:eom-R33} by determining the $f_i$ and one combination of the $s_i$.

\begin{figure}[ht]
  \centering
    \includegraphics[width=.6\textwidth]{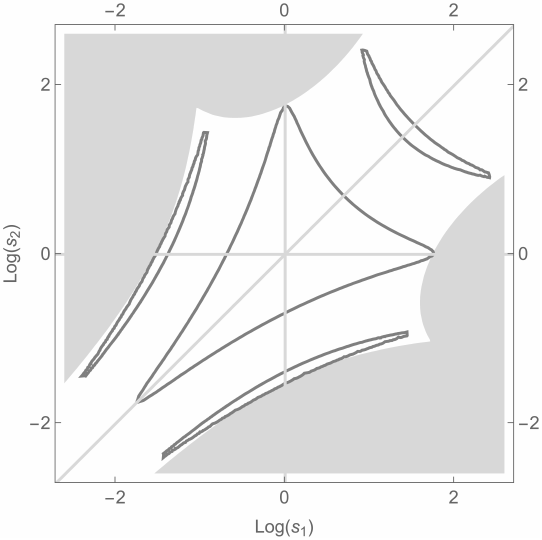}
  \caption{The gray regions are where $f_{1,2}^2\ge 0$ are \emph{not} satisfied. The dark gray locus is where $\Sigma=0$. This is where solutions exist. The light gray lines $\{s_1=1\}$, $\{s_2=1\}$, $\{s_1=s_2\}$ are fixed loci of the symmetries \eqref{eq:symflag}.}
  \label{fig:constraint}
\end{figure}

We can visualise the constraint $\Sigma=0$ as a contour in the $(s_1,s_2)$-plane. The result is shown in Fig.~\ref{fig:constraint}. The additional inequalities $f_{1,2}^2\ge 0$, necessary for $f_{1,2}$ to be real, are excluded in the gray region and are in fact implied by the constraint $\Sigma=0$. 
The permutation symmetry in the metric Ansatz \eqref{eq:ds2-F6} appeared to be mildly broken by the inclusion of $R=\alpha_3$ in \eqref{eq:G} (see footnote there), but is in fact still present: the two generators $(\alpha_1, \alpha_2, \alpha_3)\buildrel{(12)}\over\mapsto (\alpha_2, \alpha_1, \alpha_3)$, $\buildrel{(123)}\over\mapsto (\alpha_2, \alpha_3, \alpha_1)$ of the permutation group with three elements become 
\begin{subequations}\label{eq:symflag}
\begin{align}
	&(s_1, s_2, R) \buildrel{(12)}\over\mapsto 
    (s_2, s_1, R)\,,& &
    (f_1,f_2) \buildrel{(12)}\over\mapsto (f_2,f_1)\,;\\ 
    	& (s_1, s_2, R) \buildrel{(123)}\over\mapsto 
    \left(\frac{s_2}{s_1}, \frac1{s_1}, \frac R{\sqrt{s_1}}\right)\,, &  & 
    (f_1,f_2) \buildrel{(123)}\over\mapsto(f_2 s_1, -(f_1+f_2 )s_1 )\,.
\end{align} 
\end{subequations}

 The locus $\{\Sigma=0\}$ has four components: a single \emph{curved triangle}, which is left invariant by \eqref{eq:symflag}, and three \emph{boomerangs}, which are permuted by it. Two of them are in the small $s_{1,2}$ region and would be tiny; we use a logarithmic plot to make them more visible. A fundamental region for \eqref{eq:symflag} is $\{ s_1\geq s_2, s_2\geq 1 \}$, and we can limit our attention to it.

Some particular cases are worth highlighting:
\begin{itemize}
	\item As we saw below \eqref{eq:par}, $F_6$ is K\"ahler when $\alpha_3^2=\alpha_1^2+\alpha_2^2$, or $\alpha_2^2=\alpha_1^2+\alpha_3^2$, or $\alpha_1^2=\alpha_2^2+\alpha_3^2$. In these cases $\Sigma$ simplifies dramatically, leading to $\alpha_1=\alpha_2$, or $\alpha_1=\alpha_3$, or $\alpha_2=\alpha_3$. In the $(s_1,s_2)$ plane, these are the points\footnote{In Fig.~\ref{fig:constraint}, these are the intersections closest to the origin of the three boomerangs with the light gray lines.}
    \begin{equation}\label{eq:KE}
        (2,2)\,,\qquad (1/2,1)\,,\qquad (1,1/2)\,.
    \end{equation}
    In all these cases the space is actually K\"ahler--Einstein, and we reduce to the solution in \cite{Dolan:1984hv,Pope:1988xj,Martin:2008pf}.
	\item Below \eqref{eq:par} we also briefly commented about a correspondence to $\mathbb{CP}^3$. We can obtain solutions on that space by taking $s_1=s_2\equiv s$, and setting $f_1=f_2$ to take into account that only one two-cycle is present there. This gives $s=2$, $4$. The former is K\"ahler, and in fact K\"ahler--Einstein, so it is again a particular case of the solution in \cite{Dolan:1984hv,Pope:1988xj,Martin:2008pf}; the latter was found in \cite{Imaanpur:2012jx}. We will see that this value is far from the M5 instability locus, so it is not too useful for our present purposes. The $s=2$ KE solution was found to suffer from a bubble of nothing \cite{Ooguri:2017njy}; it would be interesting to see if it persists for $s=4$ as well, and for our $F_6$ solutions.
\end{itemize}

\subparagraph{Flux quantisation.} 	
So far, two parameters have not been fixed by the equations of motion: $R$, and one combination of the $s_i$, a coordinate along the $\{ \Sigma=0\}$ locus in Fig.~\ref{fig:constraint}. We will now see that flux quantisation will fix these two.

We noted in the beginning of this section that the flux parameters $f_1$ and $f_2$ do not represent the proper quantised flux parameters. 
As discussed in Sec.~\ref{sub:geo}, there are two independent four-cycles in the flag geometry, which we called $D_{13}$, $D_{23}$ around \eqref{eq:int-D13-j}. 
We then have two flux quantisation conditions for the quantised fluxes $n_1$ and $n_2$:
\be  \label{eq:fq}
\begin{split}
n_1&=\frac{1}{(2\pi \ell_P)^3}\int_{D_{13}}G
= \frac{R^2}{16\pi \ell_P^3} \left( \frac{\alpha_2^2\alpha_3^2}{\alpha_1^2}f_1+\frac{\alpha_1^2\alpha_2^2}{\alpha_3^2}(f_1+f_2)\right)\,,\\
n_2&=\frac{1}{(2\pi \ell_P)^3}\int_{D_{23}}G
= \frac{R^2}{16\pi \ell_P^3} \left( \frac{\alpha_1^2\alpha_3^2}{\alpha_2^2}f_2+\frac{\alpha_1^2\alpha_2^2}{\alpha_3^2}(f_1+f_2)\right)\, , 
\end{split}
\ee
with $\ell_P$ the eleven-dimensional Planck length. This can be inverted to determine $f_1$ and $f_2$ in terms of $n_1$, $n_2$ and the scalars $R$, $s_1$ and $s_2$. The ratio of the flux quanta reads
\be \label{eq:n12}
\frac{n_1}{n_2}=\frac{\frac{\alpha_2^2\alpha_3^2}{\alpha_1^2}\frac{f_1}{f_2}+\frac{\alpha_1^2\alpha_2^2}{\alpha_3^2}\left(1+\frac{f_1}{f_2}\right)}{\frac{\alpha_1^2\alpha_3^2}{\alpha_2^2}+\frac{\alpha_1^2\alpha_2^2}{\alpha_3^2}\left(1+\frac{f_1}{f_2}\right)}\equiv n_{12}(s)\,.
\ee
Upon using \eqref{eq:fi-sol}, the right-hand side $ n_{12}(s)$ is a function of the $s_i$ alone. Recalling the constraint $\Sigma=0$, we see that the ratio $n_1/n_2$ determines the $s_i$. We can then use one of the  \eqref{eq:fq} to determine then the overall internal scale $R$, and the AdS radius $L$ via \eqref{eq:eom-L}: they are of the form
\begin{equation}\label{eq:RL-fq}
    \frac {R^3} {\ell_P^3}= n_1 \,r_1(s_i) = n_2 \,r_2(s_i)\,,
    \qquad
    \frac {L^3} {\ell_P^3}= n_1 \,l_1(s_i) = n_2 \,l_2(s_i)\,,
\end{equation}
where $r_2/r_1=l_2/l_1=n_{12}(s)$.

Given a $(s_1,s_2)$ that satisfies \eqref{eq:n12} and $\Sigma=0$, one can rescale $n_i \to N n_i$ for a large $N$, to achieve arbitrarily large $R$ and $L$. This achieves control over these solutions, as is familiar from other flux compactifications. 


\subsection{M5 instability} 
\label{sub:m5}

\subparagraph{Minimal volumes.} We now discuss the minimal volume subspaces in each homology class.
We introduced earlier the submanifolds $S_{1,2,3}$. Each of these has an involution that has it as its fixed locus, and reverses the transverse directions. (For $S_1$, this is induced by the involution of $T_e F_6$ that reverses sign to all the non-zero roots other than $\pm r_1$.) This implies that they are extrema of the volume functional. By \eqref{eq:ds2-F6}, \eqref{eq:intSj}, 
\begin{equation}
	\mathrm{Vol}(S_a)=\pi \alpha_a^2\,.
\end{equation}  
However, the $S_a$ cannot be global minima for all values of the $\alpha_a$. 
Indeed, $S_1$ is homologous to $-S_2-S_3$; when $\alpha_1^2 \ge \alpha_2^2+ \alpha_3^2$, the union of $S_2$ and $S_3$ has smaller volume than $S_1$ by itself. On the other hand, for $\alpha_1^2 \le \alpha_2^2+ \alpha_3^2$ it is possible that $S_1$ is a global minimum in its class. Similar issues present themselves for other homology classes.

To reach more solid conclusions, we will use the theory of \emph{calibrations}. Recall that a calibration is a form $\omega$ with two properties:\\ i) at every point, on every multi-vector $V\equiv v_1 \wedge \ldots \wedge v_k$, $\iota_{V} \omega \le \sqrt{g|_V}$;\\ ii) it is closed, $\dd \omega=0$.

    A submanifold is said to be calibrated if the inequality is saturated on all its tangent spaces. A calibrated submanifold $M$ minimises the volume in its homology class: indeed, for any other $M'$ in the same homology class, $M'= M + \partial N$; using the calibration condition and Stokes' theorem:
\begin{equation}\label{eq:cal}
	\mathrm{Vol}(M)=\int_M \omega = \int_{M'} \omega +\int_N \dd \omega = \int_{M'} \omega \le \mathrm{Vol}(M')\,.
\end{equation} 

This technique appears naturally for supersymmetric solutions. It was shown in \cite{Menet:2025nbf} that it is also relevant for some classes of supersymmetry-breaking solutions in type II. However, the focus there was on trying to obtain a single calibration for a given solution. For the present eleven-dimensional solutions this strategy would not work very well: generically there our solutions do not single out a particular calibration. (For example, we noted below \eqref{eq:par} that generically the space is not K\"ahler: none of the two-forms $J$ that would calibrate the metric are closed.)

However, we can tweak the technique by introducing \emph{several} calibrations, one for each homology class of interest. The following strategy is inspired by \cite[Thm.~2.7]{LeKhongVan}. Consider for example $S_1$. A form that would automatically calibrate it is $\omega = \alpha_1^2 j_1 + \beta_2 j_2 + \beta_3 j_3$. It actually is a calibration if $\beta_2+\beta_3 = - \alpha_1^2$, $|\beta_2|\le	\alpha_2^2$, $|\beta_3|\le \alpha_3^2$. Considering the various cases for the signs of $\beta_{2,3}$, one finds that these conditions are satisfied for $\alpha_1^2 \le \alpha_2^2 + \alpha_3^2$. Thus, when this condition is satisfied, $S_1$ is actually guaranteed to minimise the volume within its homology class by \eqref{eq:cal}. When $\alpha_1^2> \alpha_2^2 + \alpha_3^2$, we can use the alternative calibration $\omega=(\alpha_2^2 + \alpha_3^2) j_1 - \alpha_2^2 j_2 - \alpha_3^2 j_3$ to show that it is the union of $S_2$ and $S_3$ that attains the minimum.

We can make this method a bit more systematic, to show that there are no other independent two-cycles in any homology class. Consider 
\begin{equation}
	\omega= \sum_{i=1}^3 \beta_a j_a\,.
\end{equation}
This is a calibration if
\begin{equation}\label{eq:cal-cond}
	\sum_a \beta_a =0 \, ,\qquad |\beta_a|\le \alpha_a^2\,.
\end{equation}
The most general homology class can be written as $S= p_1 S_1 + p_2 S_2$. Now $\int_S \omega = \pi(\beta_1 p_1 + \beta_2 p_2)$. For a given $(p_1,p_2)$, we want to maximise this as a function of the $\beta_a$, subject to \eqref{eq:cal-cond}. At this maximum, some of the inequalities will be saturated, and thus there will be a calibrated representative of the homology cycle.

Explicitly, the inequalities in \eqref{eq:cal-cond} describe a polygon $P$ in the space $(\beta_1, \beta_2)$. 
\begin{itemize}
	\item In the ``balanced'' case $\alpha_1^2 < \alpha_2^2 + \alpha_3^2$, $\alpha_2^2< \alpha_1^2 + \alpha_3^2$, $\alpha_3^2< \alpha_1^2 + \alpha_2^2$, $P$ is a hexagon with vertices $v_1=(\beta_1=\alpha_1^2,\beta_2=\alpha_3^2- \alpha_1^2)$, $v_2=(\alpha_3^2-\alpha_2^2,\alpha_2^2)$, $v_3=(-\alpha_1^2,\alpha_2^2)$, $v_4=-v_1$, $v_5=-v_2$, $v_6=-v_3$. So the maximum is 
	\begin{equation}\label{eq:hexagon}
		\mathrm{max}_{\beta_a \in P}\int_S \omega = \pi\,\mathrm{max}\{|\alpha_1^2 (p_1-p_2)+ \alpha_3^2 p_2|, |\alpha_2^2(p_2-p_1)+ \alpha_3^2 p_1|, |-\alpha_1^2 p_1 + \alpha_2^2 p_2|\}\,. 
	\end{equation}
    For example, when the maximum is realised at $v_1$, among \eqref{eq:cal-cond} we are saturating $\beta_1=\alpha_1^2$, $\beta_3(=-\beta_1-\beta_2)=-\alpha_3^2$; the cycles $S_1$, $-S_3$ are calibrated; the cycle of maximum volume in $S$ is the sum of $(p_1-p_2)$ copies of $S_1$ and $p_2$ copies of $S_3$, whose total is indeed $\alpha_1^2 (p_1-p_2)+ \alpha_3^2 p_2$, the first possibility in \eqref{eq:hexagon}. The other cases are similar.
    
	\item When $\alpha_3^2> \alpha_1^2+\alpha_2^2$, the hexagon degenerates to a rectangle with vertices $v_1'=(\alpha_1^2, \alpha_2^2)$, $v_3$, $v_4'=-v_1'$, $v_6$. The maximum is $\max\{ |\alpha_1^2 p_1+ \alpha_2^2p_2|, |-\alpha_1^2 p_1+ \alpha_2^2p_2|\}$.
	\item When $\alpha_2^2> \alpha_1^2+\alpha_3^2$, $P$ becomes a parallelogram with vertices $v_1$, $v_2'=(-\alpha_1^2, \alpha_3^2+\alpha_1^2)$, $v_4$, $v_5'=-v_2'$. The maximum is $\max\{ |\alpha_1^2 (p_1-p_2)+ \alpha_3^2 p_2|, |\alpha_1^2 (-p_1+p_2)+ \alpha_3^2p_2|\}$.
	\item Finally, when $\alpha_1^2> \alpha_2^2+\alpha_3^2$, $P$ is another parallelogram, with vertices $v_1''=(\alpha_2^2+\alpha_3^2, -\alpha_2^2)$, $v_2$, $v_3'=-v_1'$, $v_5=-v_2$. The maximum is $\max\{ |\alpha_2^2 (p_1-p_2)+ \alpha_3^2p_1|, |\alpha_2^2 (p_2-p_1)+ \alpha_3^2p_1|\}$.
\end{itemize}
In each case and for all possible maxima, 
$\mathrm{max}\int_S \omega$ is a linear combination of the volumes of the three $S_i$, just like in the example below \eqref{eq:hexagon}. This proves that there are no other independent two-cycles with a lower volume in any homology class.

Returning then to the $S_i$, for each value of the triple $\alpha_i$ we have seen that there are only two possibilities: either the sphere $S_i$ is the lowest volume two-cycle, or it is the union of the remaining two other spheres that realises the global minimum. 

\begin{figure}[!ht]
  \centering
    \includegraphics[width=.55\textwidth]{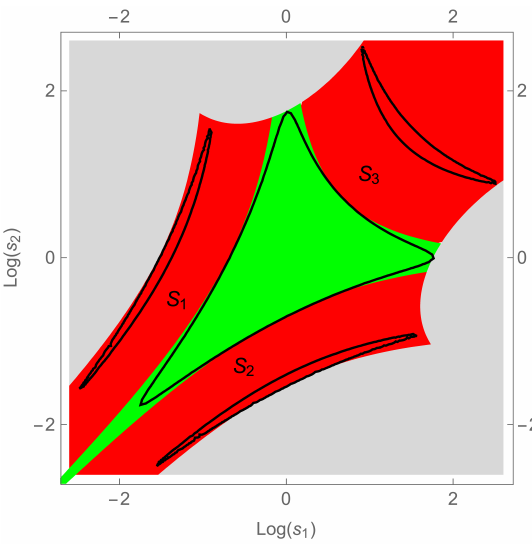}
  \caption{The different regions with different minimal volume representatives in the $(s_1,s_2)$ plane. In the green region, each sphere is the global minimum two-cycle. In each red region denoted $S_i$, the $S_i$ sphere has a higher volume than the union of the remaining two spheres. In these regions, the other spheres minimise the volume.}
  \label{fig:stab plane}
\end{figure}

We can now check which of these cases is realised in the $(s_1,s_2)$ plane. This is done in Fig.~\ref{fig:stab plane}.
For the first region of interest in the $(s_1,s_2)$ plane, i.e.~the curved triangle, we see from Fig.~\ref{fig:stab plane} that each sphere is the global minimum within its homology class. This is the balanced case mentioned above ($\alpha_1^2 < \alpha_2^2 + \alpha_3^2$, $\alpha_2^2< \alpha_1^2 + \alpha_3^2$, $\alpha_3^2< \alpha_1^2 + \alpha_2^2$). For the upper-right boomerang we see, again from Fig.~\ref{fig:stab plane}, that $S_1$ and $S_2$ each minimise the volume of their homology class again, while $S_3$ has a higher volume than the union of $S_1$ and $S_2$.

\subparagraph{M5-brane nucleation.}

Now that we have identified the minimal volume representatives in each homology class, for all of our solutions, we can consider M5-branes wrapping these two-dimensional internal cycles. The corresponding M5-branes will therefore be domain walls in AdS$_5$. 

\begin{figure}[!ht]
\centering
\begin{minipage}[c]{0.45\textwidth}
    \centering
    \includegraphics[width=\textwidth]{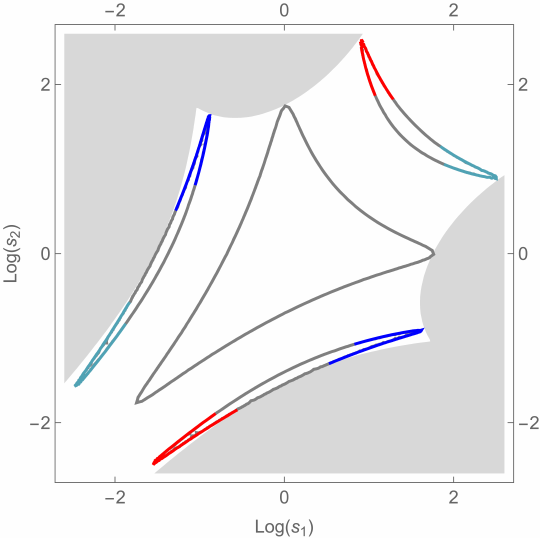}
\end{minipage}
\hfill
\begin{minipage}[c]{0.50\textwidth}
    \centering
    \includegraphics[width=\textwidth]{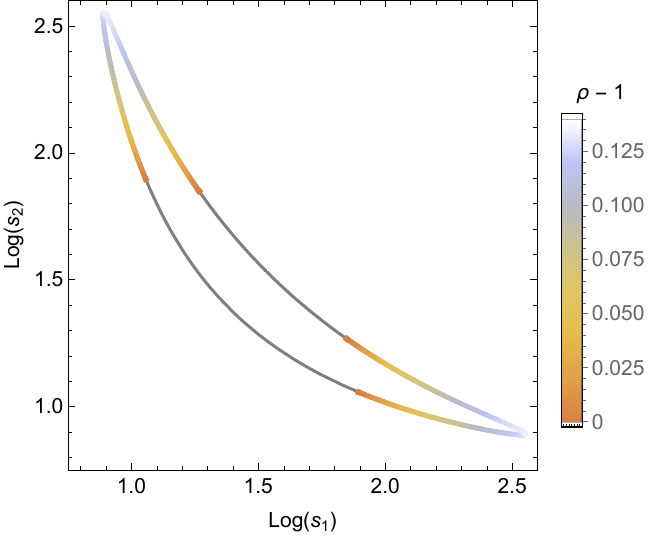}
\end{minipage}
\caption{Left: the regions where $S_1$, $S_2$, $S_3$ are unstable are shown in red, green, blue, respectively. Right: a zoom on the upper-right boomerang; the stability ratio $\rho$ for $S_1$ ($S_2$) in its upper-left (lower-right) corner. Only the unstable solutions have been colour-coded, the stable ones are represented in grey.}
\label{fig:flag ratios}
\end{figure}

We can evaluate the stability ratio $\rho$ in \eqref{eq:rho} for M5-branes wrapping the different spheres available within the flag geometry, using the four-form flux $G$ \eqref{eq:G} and the equations of motion \eqref{eq:eoms}. We obtain
\begin{equation}\label{eq:stability_ratios}
    \rho_{S_1}= \frac{LR^2 f_1}{4\min(\alpha_1^2,\alpha_2^2+\alpha_3^2)} \,,\qquad
    \rho_{S_2}= \frac{LR^2 f_2}{4\min(\alpha_2^2,\alpha_1^2+\alpha_3^2)} \,,\qquad
    \rho_{S_3}= \frac{LR^2 (f_1+f_2)}{4\min(\alpha_3^2,\alpha_1^2+\alpha_2^2)} \,.
\end{equation}
The results are displayed in Fig.~\ref{fig:flag ratios}. For the KE case $s_1=s_2=2$, all the stability ratios satisfy $\rho_{S_i}=\frac1{\sqrt2}$, reproducing \cite[(9)]{Martin:2008pf}; we already mentioned that these solutions are stable.

From the above discussion, we know that it makes sense for an M5-brane to wrap $S_1$ or $S_2$, be it for the curved triangle or the boomerang solutions, since these spheres are always the minimum volume representatives within their homology class. However, it does not really make sense to consider the physics of an M5-brane wrapping $S_3$ for the boomerang solutions, since for these solutions $S_3$ is not a globally stable cycle, as shown in Fig.~\ref{fig:stab plane}.


\subsection{Hierarchies}
\label{sub:pheno}

Motivated by the dark bubble scenario, we now focus on very specific flag solutions: the ones for which the M5-brane wrapping a given sphere is only very slightly super-extremal, meaning that the stability ratio is only very slightly larger than one. In the left pane of Fig.~\ref{fig:flag ratios}, these correspond to the locations on the boomerangs that are coloured and in the vicinity of the grey-coloured boundaries. For each $S_i$, up to the symmetries \eqref{eq:symflag} there are two such points, both on a boomerang. We focus on instabilities arising from an M5-brane wrapping $S_1$, whose extremal points are the red-grey boundaries on the upper-right boomerang of the left pane of Fig.~\ref{fig:flag ratios}, of which a zoomed-in version can be found in the right panel.
We denote the extremal point on the lower branch by $p_1\simeq (2.87736,6.65331)$ and the one on the upper branch by $p_2\simeq (3.55091,6.34568)$. These points do not correspond to proper solutions, as fluxes cannot be appropriately quantised there. This can be motivated as follows: the points $p_1$ and $p_2$ are solutions to the equations $\Sigma =0$ and $\rho_{S_1}=1$, where these quantities were defined by eqns.~\eqref{eq:Sigma} and~\eqref{eq:stability_ratios}. Analytically, they imply that $s_1 = \frac{s_2}{\sqrt{1-6s_2+ s_2^2}}$, and that $s_2$ is given by the two roots near $6$ of the polynomial
\be P(s_2)=-2584(1+s_2^8) - 19752 s_2(1+s_2^6) + 8689 s_2^2(1+s_2^4) - 459612 s_2^3(1+s_2^2) + 
3496182 s_2^4\,.\ee
Applying the rational root theorem to $P(s_2)$, one finds that the roots $p_1$ and $p_2$ are not rational. Moreover, flux quantisation requires that the function $n_{12}(s)$ \eqref{eq:n12} must be rational. Evaluating it on $p_{1,2}$ gives two algebraic numbers whose minimal polynomials (over $\mathbb{Q}$) have degree eight; so $n_{12}(p_1)\simeq 0.795888$ and $n_{12}(p_2)\simeq -0.891851$ are not rational either. This implies that, although $p_1$ and $p_2$ are solutions to the equations of motion, they do not correspond to proper M-theory vacua due to flux quantisation.
Nevertheless, there exist near-extremal solutions with properly quantised fluxes, and we exploit them for the dark bubble scenario. 

Intuitively, an extremal solution (if it existed) would correspond to a static bubble; so, a nearly-unstable solution should correspond to a bubble that expands slowly, i.e.~one with a small cosmological constant. We consider such solutions near $p_1$; a similar story for $p_2$ applies. To pick such a vacuum, we should select a pair $(s_{1+},\, s_{2+})$ satisfying $\Sigma=0$ (on the coloured side near $p_1$) for which the flux parameters $n_1=n_{1+}$ and $n_2=n_{2+}$ are \textit{integers} that make $n_{12}$ arbitrarily close to $n_{12}(p_1)\simeq 0.795888$. The Dirichlet approximation theorem guarantees that we can do this up to a precision $\delta^2$:
\begin{equation}
    \left|n_{12}(p_1) - \frac{n_{1+}}{n_{2+}}\right|< \delta^2\,, \qquad \delta^{-1} \sim n_{2+}\,, \qquad n_{1+}, n_{2+} \in \mathbb{Z}_0\,,
\end{equation}
with $\mathrm{gcd}(n_{1+},n_{2+})=1$. Since $n_{12}(p_1)$ is order one, we must have that $n_{1+}$ is also order $\delta^{-1}$.

For example, if we pick $\delta^2 = 10^{-10}$ we can use $n_1=16841$, $n_2=21160$. This determines a slightly unstable solution $p_+=(s_{1+},s_{2+})$. Comparing with the extremal point $p_1$, we find that $s_{i+}-s_i(p_1) \sim \delta^2$, which can be concluded from the fact that the derivative $\partial_{s_i} n_{12}$ is of order one (together with $\partial_{s_i} n_{12}\simeq \Delta n_{12}/\Delta s_i \sim 1$, and the approximation theorem above). 
After picking a slightly unstable solution with a specific choice for $n_{1+}$ and $n_{2+}$, we consider the nucleation of a single M5-brane wrapping the two-cycle $S_1$. 
Using that the derivative of the stability ratio for this M5-brane, $\partial_{s_i} \rho$, is of order one, we find that the super-extremality parameter satisfies $\epsilon \sim \delta^2$.\footnote{As we saw in Sec.~\ref{sec:rev}, $\epsilon$ is defined via $\sigma = \sigma_\mathrm{crit}(1 - \epsilon)$. In our context, the tension $\sigma$ is computed using the probe perspective; the five-dimensional perspective would not even resolve the localised brane. So it makes more sense to use the probe perspective \eqref{eq:repro-sigma} for $\sigma_\mathrm{crit}$ rather than the 5d formula \eqref{eq:sigma-crit}. (We did show that these two quantities are approximately equal, but the equality works to a precision $\delta$.)}

The M5-brane wrapping $S_1$ triggers a decay to a new vacuum with flux quanta $n_{1-}=n_{1+}-1$, $n_{2-}=n_{2+}$. Computing the cosmological constant $\Lambda_-$, we find that it is more negative than $\Lambda_+$ for the previous vacuum, as it should. But perhaps surprisingly, it is not towards the stable (grey) side, but even further on the unstable (coloured) side. This is illustrated in Fig.~\ref{fig:Vminplots}.
In other words, the new vacuum $(s_{1-},s_{2-})$ is itself non-perturbatively unstable; this means that it will nucleate new bubbles, which are slightly faster than the previous ones. It would be interesting to estimate how this affects the lifetime of our dark bubble, but we will not do so here.

The new vacuum lies much further from the extremal point than the old one, as
\begin{equation}
\frac{n_{1-}}{n_{2-}}= \frac{n_{1+}}{n_{2+}}-\frac1{n_{2+}}\sim n_{12}(p_1)-\delta\,,
\end{equation}
whereas the old vacuum at $(s_{1+},s_{2+})$ was only $\delta^2$ away from $n_{12}(p_1)$. From this, we also conclude that $|s_{i+}-s_{i_-}| \sim \delta$.

We are now ready to derive parametric estimates for the dark bubble scenario on our M5-brane.
Recalling $k=1/L$, we use eq.~\eqref{eq:RL-fq} and the estimates above to derive that
\begin{equation}
    \ell_P \Delta k \sim  n_2^{-1/3} |s_{i+}-s_{i-}| \cdot \partial_{s_i}l_2^{-1/3}\Big|_{s_{i+}} \sim \delta^{\frac{4}{3}}\,;
\end{equation}
we have checked that $\partial_{s_i}l_2^{-1/3}|_{s_{i+}}$ is order 1. Using the relation $\ell_5^3/\ell_4^2= 2k_+k_-/\Delta k$ between the 4d and 5d Planck scales, the equalities \eqref{eq:l45}, \eqref{eq:lambdaeps}, and the usual compactification relation $\ell_5^{-3}\sim \ell_P^{-9} R^6$, we can finally estimate all remaining scales:
\begin{equation}
    \ell_4\sim \ell_5 \sim \ell_P \delta^{2/3} \,,\qquad R\sim \ell_4 \delta^{-1}\,, \qquad \ell_4^4 \rho_\Lambda \sim \delta^4\,.
\end{equation}
In other words,
\begin{equation}\label{eq:Lambda-est}
    \rho_\Lambda  \sim R^{-4}\sim m_{\rm KK}^4 \,.
\end{equation}
Quite remarkably, \eqref{eq:Lambda-est} is the same type of relation invoked in \cite{Montero:2022prj} by using swampland criteria. Here it was just a consequence of our model.\footnote{On the other hand, a weakness of our derivation is the use of the probe approximation. It is conceivable that taking the back-reaction into account would impact our estimates.}
Note that we also obtain the following hierarchies:
\begin{equation}
    \ell_4 \sim \ell_5 \ll \ell_P \ll \rho_\Lambda^{-1/4}\sim R\,.
\end{equation}
Apart from the fact that the 5d Planck length is not much smaller than the 4d one, the other scales satisfy the same hierarchies as in the dark bubble model of refs.~\cite{Danielsson:2022lsl,Danielsson:2023alz}. The fact that $\ell_4$ and $\ell_5$ are parametrically the same can be traced back to the flux dependence of the AdS scale, which is different between M- and string theory.

One can go beyond these scaling behaviours and evaluate $\Lambda_4$ and $L$ for a given dark bubble. To do so, pick a given order for $\delta$, which uniquely sets $n_{1+}$ and $n_{2+}$. (For a given $\delta$, there is a unique optimal approximant $\frac{n_{1+}}{n_{2+}}$ for $n_{12}(p_1)$.) This in turn selects a starting AdS$_5$ vacuum, with an AdS radius $L_+$, and a bubble with a given deviation from extremality $\epsilon$ (which we evaluate in the probe perspective, as explained in the previous footnote). 
We can then compute $\Lambda_4$ using \eqref{eq:Lambda4-probe}. 
We have a discrete set of starting AdS$_5$ vacua, determined by the increasing order of the precision $\delta$. They each have a corresponding bubble with a given cosmological constant $\Lambda_4$, which turns out to be approximately $5000 \:\delta^2 n_2^{-2}$. For instance, an interesting dark bubble, in the vicinity of $p_2$ and for $\delta=10^{-32}$, has
\begin{equation}
\ell_4^2 \Lambda_4 = \ell_4^4\, \rho_\Lambda\simeq 8.7\times 10^{-121}\qquad L\simeq 37\, \mathrm{\mu m}\qquad R\simeq 68\, \mathrm{\mu m}\,.
\end{equation}
This is of the same order of magnitude as the observed cosmological constant, in our reduced Planck mass conventions. 

Interestingly, the size of the extra dimensions is similar to scales currently probed by short-distance gravity experiments. Other types of experiments (e.g.~neutron stars) give much more stringent constraints for a large number of extra dimensions. The dark bubble scenario leads, however, to quite different predictions from those of a standard KK compactification \cite{Danielsson:2025ffk}. The KK modes of the M5 fields would be non-gravitational and potentially more dangerous. This ultimately depends on how one would introduces matter in this model. Some potential strategies for that would be to include singularities, or to reduce to IIA and introduce D-branes. It would be interesting to consider these issues in more detail.

\subsection{Comments on perturbative stability}
\begin{figure}[ht]
    \centering
    \includegraphics[width=0.75\linewidth]{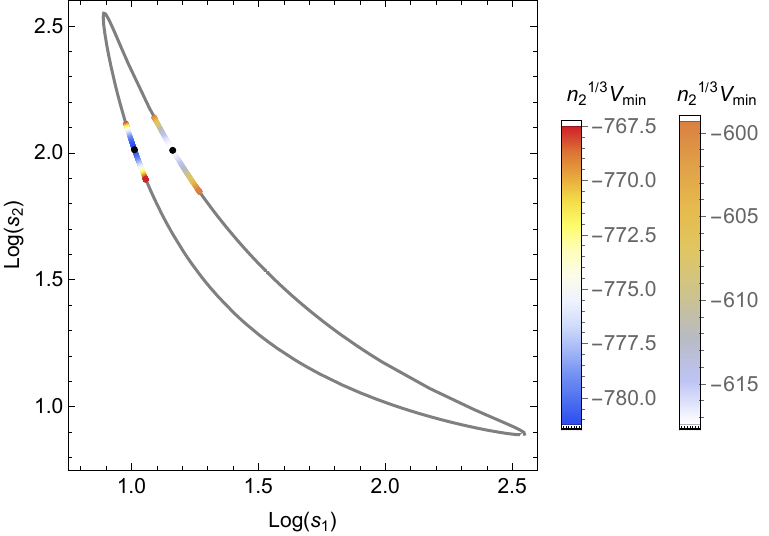}
    \caption{Plot of $n_2^{1/3}V_\mathrm{min}$ on the contour, only for $S_1$. The plots are only coloured in the regions where $V_\mathrm{min}$ takes its minimal value (indicated by the black points) and its value at the extremal point. As argued in the main text, the value of $n_{12}$ becomes minimal at black points, and one of the masses of the scalars vanishes.}
    \label{fig:Vminplots}
\end{figure}

Having obtained the AdS$_5$ solutions from the eleven-dimensional equations of motion, we now use the scalar potential from dimensional reduction to determine the masses of the three scalar modes and assess their perturbative stability. The potential, whose extrema reproduce \eqref{eq:eoms}, receives only contributions from the non-vanishing positive curvature $R_6$ of the internal manifold and the four-form flux $G$.
Performing the dimensional reduction of M-theory with these ingredients, we find the following potential:
\begin{equation}
    V = \left(-R_6 +\frac{1}{2}|G|^2\right)R^{-4}(s_1 s_2)^{\frac{2}{3}}\,,
\end{equation}
where $V$ is dimensionless (the full energy density being $M_\mathrm{Pl}^5 V$), and $R$ is expressed in 11d Planck units. The Ricci scalar $R_6$ can be obtained from the Ricci tensor components in eq.~\eqref{eq:eoms} and $|G|^2$ is easily obtained from its definition. 
Setting the gradient of $V$ (with respect to the fields $R$, $s_1$ and $s_2$) reproduces our equations of motion \eqref{eq:eoms}, provided we first replace the flux parameters $f_1$ and $f_2$ in favour of the quantised fluxes $n_1$ and $n_2$ as in \eqref{eq:fq}. 
To find the masses, we need the Hessian of the potential and the field-space metric, which can also be obtained by the dimensional reduction and is given with $\phi^i = \{s_1, s_2, R\}$ as follows:
\begin{equation}
    \frac{1}{2} g_{ij} \dd \log \phi^i \dd \log \phi^j\,, \qquad 
    g = \frac{1}{3}\begin{pmatrix}
        5 & 2  & -18 \\
        2  & 5 & -18 \\
        -18 & -18 & 108
    \end{pmatrix}\,.
\end{equation}
Focussing on the instability region of the upper-right boomerang in Fig.~\ref{fig:flag ratios} (right), we find that there are special points at which one of the scalars becomes massless. These are indicated in black in Fig.~\ref{fig:Vminplots}. Between these points and the extremal points, the masses squared are positive, whereas on the other side the extremum is a saddle point of the potential. Not all solutions satisfy the BF-bound there, in particular those close to the tip of the boomerang.

\begin{figure}
    \centering
    \includegraphics[width=0.5\linewidth]{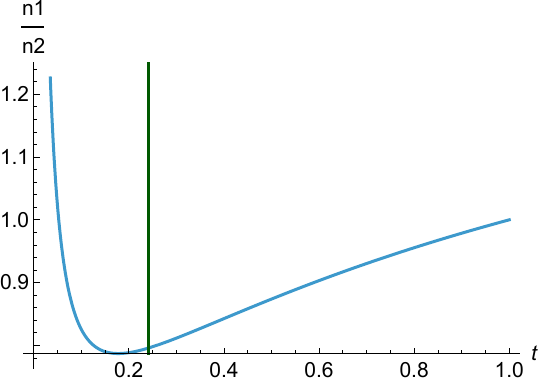}
    \caption{The quantity $n_{12} = n_1/n_2$ as a function of $t$ for the lower branch of the upper-right boomerang. The green line indicates the value of $t$ at which the extremality ratio $\rho_{S_1}$ is 1.}
    \label{fig:n12plot}
\end{figure}

Another interesting aspect of the setup is that there exist pairs of points on the boomerang that correspond to different vacua of the \textit{same} scalar potential, i.e.~with the same flux quanta $n_1$ and $n_2$. 
To see this, we can investigate the equations of motion eq.~\eqref{eq:eoms} by introducing the parameters $t = f_2/f_1$ and $r = R f_1$. One can then formally solve these equations of motion for $s_1$, $s_2$ and $r$ as a function of the parameter $t$. This gives four real solutions for positive $r$, and making a parametric plot in $t$ gives the same as Fig.~\ref{fig:constraint}. One of these four solutions constructs the entire upper-right boomerang of figure Fig.~\ref{fig:constraint}, with positive values of $t$ resulting in the lower curve, and negative values in the upper curve.\footnote{The upper curve can also be constructed by considering one of three other solutions and apply the symmetry transformations \eqref{eq:symflag}, such that one can work with positive $t$ for this branch as well.}
For the upper-right boomerang, we can plot $n_{12}$ of \eqref{eq:n12} as a function of $t$ (see Fig.~\ref{fig:n12plot}), and we see that there indeed exist different vacuum solutions with different $t$ that have the same value of $n_{12}$, as the latter obtains a minimum. 
This minimum corresponds precisely to the point in our flux landscape where the value of the potential has a minimum. To see this, we note that we can write the solution to the equations of motion as $(s_1(t), s_2(t), r(t))$. So the value at the extremum of the potential is given as follows:
\begin{equation}
    V_\mathrm{min}(t) = V\!\left(s_1(t), s_2(t), r(t)\,; \,n_{12}(t)\right).
\end{equation}
We emphasise that $n_{12}$ varies but that we keep $n_2$ fixed.
Extremising this w.r.t. $t$ gives
\begin{equation}
    \frac{\dd V_\mathrm{min}}{\dd t} = (\partial_{\phi^i} V )\dot{\phi}^i +(\partial_{n_{12}}V)  \frac{\dd n_{12}}{\dd t} = (\partial_{n_{12}}V) \frac{\dd n_{12}}{\dd t}\,,
\end{equation}
where $\phi^i = (s_1, s_2, r)$ and $\dot{\phi}^i = \dd \phi^i / \dd t$. The first term vanishes by virtue of the equations of motion, i.e. $\partial_{\phi^i} V =0 $. So we see that $V_\mathrm{min}(t)$ reaches an extremum $V_\mathrm{min}^*$ when $n_{12}(t)$ also reaches an extremum, say at $t^*$. We checked that for the slightly super-extremal M5-brane solutions, the extrema are minima, and are illustrated in Fig.~\ref{fig:Vminplots}.
Moreover, a similar argument shows that these points correspond precisely to where one of the masses vanishes. To see this, we evaluate the three field derivatives of $V$ on the curve of the solution, which gives
\begin{equation}
    \partial_{\phi^i}V\left(s_1(t), s_2(t), r(t)\,;\,n_{12}(t)\right) =0\,,
\end{equation}
Taking a derivative with respect to $t$ of these three equations gives
\begin{equation}
    (\partial_{\phi^i}\partial_{\phi^j}V) \dot{\phi}^j +\partial_{n_{12}}\partial_{\phi^i}V \frac{\dd n_{12}}{\dd t}=0\,,
\end{equation}
At the extremum of $n_{12}$ (with $t=t^*$), we see that the second term vanishes, and it implies that $\dot{\phi}^j(t^*)$ (when non-vanishing) is an eigenvector of the Hessian of the potential with eigenvalue zero. So the determinant of the Hessian vanishes at this point, and hence so does the determinant of the mass matrix, indicating that one of the masses must vanish.

We end this section by commenting that a more complete study of the perturbative spectrum would require carrying out a full KK reduction. In general this is a notoriously complicated task. As an encouraging sign, we note that for AdS$_5\times {\rm KE}_6$ the problem of full perturbative stability reduces to the spectrum of the Laplacian on $(1,1)$-forms \cite{Martin:2008pf}; for $\mathbb{CP}^3$ stability does hold, but in our $F_6$ case with \eqref{eq:KE} the answer is not currently known. It would be interesting to extend this computation to our more general solutions. 

\section{Uplifts from seven dimensions} 
\label{sec:7d}

We now present a class of supersymmetry-breaking solutions of the form AdS$_5 \times S^2$ in gauged maximal $d=7$ supergravity. These can be uplifted to eleven-dimensional supergravity. Unfortunately, we will find them all to be perturbatively unstable.

These supergravity solutions emerge as part of a larger class where $S^2$ can be replaced by any Riemann surface $\Sigma_g$. They are the supersymmetry-breaking counterparts of the very well-known MN \cite[Sec.~4.1]{Maldacena:2000mw} and BBBW \cite{Bah:2012dg} supersymmetric vacua.\footnote{As we will see, a couple of our supersymmetry-breaking solutions were found in \cite[Sec.~3]{Gauntlett:2002rv}.} There is also a variant in minimal gauged $d=7$ supergravity, which can be uplifted to $d=11$ or to massive IIA. However, the moduli of $\Sigma_{g\ge 1}$ are expected to acquire a potential from stringy corrections, stabilising them to values that might correspond to singularities. Since we are not able to control this process, these more general solutions are not under sufficient control in string theory in their present incarnation.\footnote{A more obvious, less vast and equally problematic set of solutions would be obtained by simply replacing AdS$_7$ with AdS$_5\times \Sigma_g$ in known solutions of M-theory and massive IIA.} For this reason, we will focus on the $\Sigma=S^2$ case throughout our presentation.

\subsection{Maximal supergravity solutions} 
\label{sub:7dsol}

The bosonic fields of maximal gauged $d=7$ supergravity are \cite{Pernici:1984xx}: a metric; an SO(5) gauge field; five three-forms; 14 scalars parametrising the coset $\mathrm{SL}(5)/\mathrm{SO}(5)$. The fermions are four gravitini, and sixteen dilatini.

\subparagraph{7d solutions.} As in \cite{Cvetic:1999xp,Liu:1999ai}, we consider an abelian Ansatz, with scalars $X_i \equiv e^{2\lambda_i}$; gauge fields $A_i$, $i=1,2$,  in the Cartan subalgebra $\mathfrak{u}(1) \oplus \mathfrak{u}(1)$, with field strengths $F_{2\,i}=\dd A_i$; and a single three-form $C$, with field strength $F_4$. For the action and full equations of motion, we follow \cite[(2.5--10)]{Liu:1999ai}.\footnote{They use a notation $F_{\mu\nu}^2\equiv F_{\mu\nu} F^{\mu\nu}$ in the action, and $F^2\equiv F_{\mu\nu} F^{\mu\nu}$, $F_{\mu\nu}^2\equiv F_{\mu\rho} F^{\nu\rho}$ in the equations of motion.} 

We then use the Ansatz
\begin{equation}\label{eq:7dmax-ans}
\begin{split}
	\dd s^2_7 &= L^2\,\dd s^2_{\mathrm{AdS}_5}+R^2\,\dd s^2_\Sigma \, ,\qquad
	X_i=\text{const.}\,,\\ 
	F_4&=0\,,\qquad
	F_{2\,,i}= f_i R^2 \mathrm{vol}_\Sigma\,.
\end{split}
\end{equation}
For now $\Sigma$ can be any compact Riemann surface, but as we explained at the beginning of this section, we are especially interested in the case $\Sigma=S^2$.

The gauge-field and three-form equations of motion are now identically satisfied. The scalar and gravity equations reduce to:
\begin{equation}\label{eq:7dmax-eom}
\begin{split}
	f_1^2 &= \frac{m^2}{2X_1^2 X_2^4} (1-X_1^3 X_2^2)(2X_1^2 X_2^3-1)\,,\\
	f_2^2 &= \frac{m^2}{2X_1^4 X_2^2} (1-X_1^2 X_2^3)(2X_1^3 X_2^2-1)\,,\\
	L^2 &= \frac8 {m^2} \frac{X_1^4 X_2^4 }{2X_1^3 X_2^2+2X_1^2 X_2^3-1}\,,\\
	\kappa &=  -m^{2}R^2 X_1^{-4} X_2^{-4} \left(\frac12 + (2 X_1^3 X_2^2-1)(2 X_1^2 X_2^3-1)\right)\,,	
\end{split}	
\end{equation}
where $\kappa=1$, $0$, $-1$ correspond to $S^2$, $T^2$, $\Sigma_{g\ge 2}$ as usual.
Non-negativity of $f_i^2$ imposes
\begin{equation}\label{eq:X-regions}
	u\equiv X_1^3 X_2^2,\,\quad v\equiv X_1^2 X_2^3 \ \ \in \ \ 
		[0,1/2]\times [1,\infty) \ \ \cup \ \ [1/2,1]^2 \ \ \cup \ \ [1,\infty)\times [0,1/2]\,. 
\end{equation}
$L^2\ge 0$ is then automatically achieved. Positivity of $R^2$ is obtained for $\kappa=1$ ($-1$) in the green (red) regions of Fig.~\ref{fig:7d-all}, with $\kappa=0$ allowed at the boundary between the two.

\subparagraph{Supersymmetry.} Our Ansatz includes that of \cite{Bah:2012dg}, which leads to supersymmetric solutions. Using (3.8) in that reference, we find that it is the locus
\begin{equation}\label{eq:bbbw}
	u+v=3/2\,.
\end{equation}
This is shown as the gray line in Fig.~\ref{fig:7d-all}. This locus also includes the MN2 solutions \cite[Sec.~4.1]{Maldacena:2000mw} for  $X_1=2^{-3/5}$, $X_2=2^{2/5}$, $u=1/2$, $v=1$ (or $X_1\leftrightarrow X_2$, $u\leftrightarrow v$) and MN1 \cite[Sec.~4.2]{Maldacena:2000mw} $X_1=X_2=3^{1/5}2^{-1/5}$, $u=v=3/4$.

\subparagraph{Uplift.} These solutions uplift to $d=11$ supergravity. Upon using \cite[(4.1,2)]{Cvetic:1999xp} (with $g_{\text{\tiny\cite{Cvetic:1999xp}}}=m$, $F^i_{\text{\tiny\cite{Cvetic:1999xp}}}=2F^i$), one obtains
\begin{equation}
\begin{split}\label{eq:uplift}
\dd s^2_{11}&= \tilde \Delta^{1/3} \dd s^2_7 + m^{-2} \tilde \Delta^{-2/3} \left(X_0^{-1} \dd \mu_0^2 + \sum_{i=1}^2 X_i^{-1} (\dd \mu_i^2 + \mu_i^2 D \phi_i^2)\right) \,,\\
 *G &= m \left(2 \sum_{\alpha=0}^2 (X_\alpha^2 \mu_\alpha^2- \tilde \Delta X_\alpha)+ \tilde \Delta X_0\right)\mathrm{vol}_7 + \frac1{m^2} \sum_{i=1}^2 f_i X_i^{-2} L^5 \dd(\mu_i^2) \wedge D \phi_i \wedge {\rm vol}_{\rm AdS_5}\,,
\end{split}
\end{equation}
where $\phi_i\in[0,2\pi]$, $D \phi_i \equiv \dd \phi_i + 2 m F_i$, $F_i= f_i R^2 a$, $\dd a={\rm vol}_\Sigma$; $X_0 \equiv X_1^{-2} X_2^{-2}$, $\tilde \Delta = \sum_{\alpha=0}^2 X_\alpha \mu_\alpha^2$; and the $\mu_\alpha$ are taken such that $\sum_{\alpha=0}^2 \mu_\alpha^2=1$. A common choice is $\mu_0 = \cos \theta \cos \psi$, $\mu_2 = \cos \theta \sin \psi$, $\mu_1 = \sin \theta$, with ranges $\theta\in [0,\pi/2]$, $\psi\in[0,\pi]$. With this choice, we can also write $\tilde \Delta= X_1^{-2}X_2^{-2}(u \sin^2\theta + \cos^2\theta(\cos^2\psi+ v \sin^2\psi))$. The internal topology is that of an $S^4$-fibration over $\Sigma_g$. 

We can also compute the Hodge star to obtain a possible three-form potential:
\begin{equation}\label{eq:C3}
G= \dd C_3 \,,\quad C_3= \frac1{m^3} (c_1 \wedge D\phi_1+2mf_1 R^2 \mu_0 {\rm vol}_\Sigma) \wedge D\phi_2 \,,\quad c_1 \equiv \frac{(uv)^{2/5}}{\tilde \Delta} \left( u\mu_1^2 \dd \mu_0 -\frac12 \mu_0 \dd(\mu_1^2)\right)\,.
\end{equation}

\subparagraph{Stability ratio.} The only two-cycle that an M5 bubble can wrap is $\Sigma(=S^2)$. A subtlety is the presence of the angles $\phi_i$ in \eqref{eq:uplift}: each is a non-trivial $S^1$-fibration over $S^2$, and thus has no zero section. The way out is to go to the locus where they shrink: $\mu_1=\mu_2=0$, or in other words $\theta=\psi=0$.  
Using again \eqref{eq:uplift}, the tension and charge density \eqref{eq:fq} read
\begin{equation}\label{eq:sq-7d}
\begin{split}
	\sigma&= 4\pi T_{\mathrm{M5}} R^2 \tilde \Delta|_{\theta=\psi=0} =4\pi T_{\mathrm{M5}} R^2 X_1^{-2} X_2^{-2} \, ,\\
	q h &= 4\pi T_{\mathrm{M5}} m L R^2 \left(2 \sum_{\alpha=0}^2 (X_\alpha^2 \mu_\alpha^2- \tilde \Delta X_\alpha)+ \tilde \Delta X_0\right)|_{\theta=\psi=0} \\
	&=4\pi T_{\mathrm{M5}} m L R^2 X_1^{-4} X_2^{-4} (2 X_1^3 X_2^2+ 2 X_1^2 X_2^3 -1)\,.
\end{split}
\end{equation}
The stability ratio \eqref{eq:rho} is then simply
\begin{equation}\label{eq:rho-7d}
	\rho =\sqrt{u+v-1/2}\,. 
\end{equation}
As expected, we see that $\rho=1$ on the supersymmetric locus \eqref{eq:bbbw}.

\subparagraph{Minimal supergravity.}

A variant of the solution in this subsection can be obtained by considering \emph{minimal} gauged $d=7$ supergravity. Its bosonic fields are the metric, an SU(2) gauge field $F_2^\alpha$, $\alpha=1,2,3$, and a scalar $X$, with a coupling constant $g_0$ \cite{Townsend:1983kk}. We can consider an Ansatz as in \eqref{eq:7dmax-ans}, with $X=$ const. and $F_2^\alpha = f \delta^\alpha_3 R^2 {\rm vol}_\Sigma$. With the equations of motion \cite[(4.17--19)]{Passias:2015gya}, the solutions are given by \eqref{eq:7dmax-ans} with $m^2=g_0^2/2$, $X_i=X$, $f_i = 2^{-3/2} f$. These now exist only for $\kappa=-1$, $X\in (2^{-1/5},1)$.\footnote{The values of these two endpoints are those of the supersymmetry-breaking and supersymmetric AdS$_7$ vacua of minimal 7d supergravity, respectively.} They can again be uplifted to M-theory \cite{Lu:1999bc}, but also to massive IIA in infinitely many different ways \cite{Passias:2015gya}, one for each supersymmetric AdS$_7$ solution in that theory. However, $\kappa=-1$ means that the moduli problem is unavoidable in this case.

\subsection{Flux quantisation} 
\label{sub:7d-fq}

Flux quantisation for the four-form, $\mathbb{Z} \ni N\equiv (2\pi \ell_P)^{-3}\int_{S^4}G$, can be worked out easily using $C_3$ from \eqref{eq:C3}. The only four-cycle is the $S^4$ fibre; integrating on it we obtain $m^{-3}= \pi N \ell_P^3$. This is indeed compatible with the prefactor in the MN2 solution (see \cite[(2.7)]{Gaiotto:2009gz}).

\begin{figure}[ht]
  \centering
	\begin{minipage}[c]{0.45\textwidth}
	    \centering
	    \includegraphics[width=\textwidth]{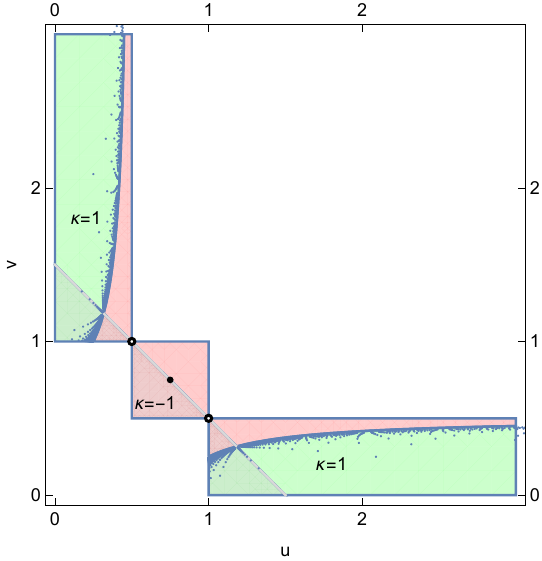}
	\end{minipage}
	\hfill
	\begin{minipage}[c]{0.4\textwidth}
	    \centering
	    \includegraphics[width=\textwidth]{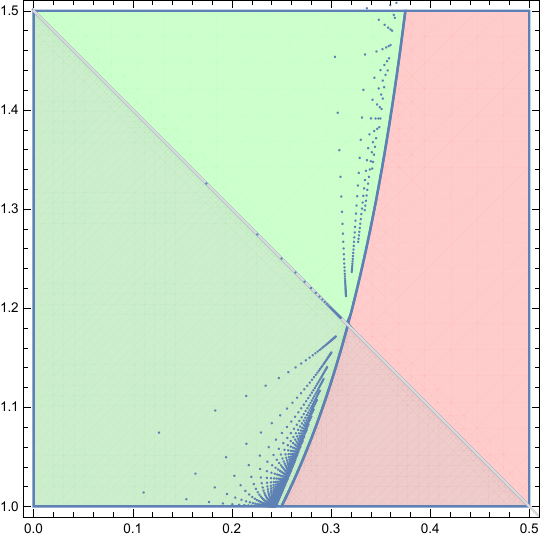}
	\end{minipage}
  \caption{Left: the allowed regions in \eqref{eq:X-regions}. The green (red) region represents $\kappa=1$ ($-1$). The gray diagonal line represents the BBBW supersymmetric solutions. The darker (lighter) regions are where the M5 is stable (unstable). The dots in the green region are allowed by flux quantisation. They accumulate at the green-red boundary. Right: a zoom on the ``fan'' structure in \eqref{eq:rays}.}
  \label{fig:7d-all}
\end{figure}

Flux quantisation for the two-forms $F_i$ uplifts to that of Chern classes for the fibrations of the coordinates $\phi_i$. We need to impose
\begin{equation}
	\mathbb{Z} \ni p_i \equiv \frac1{2\pi}\int_\Sigma 2m F_i = 4 m R^2 f_i (g-1)\,.
\end{equation}
(We used $\int_\Sigma \mathrm{vol}_\mathrm{S}= 2\pi |\chi|= 4\pi |g-1|$.) Using \eqref{eq:7dmax-eom}:\footnote{The signs of the $f_i$ are immaterial in \eqref{eq:7dmax-eom}; we have fixed them here so that $p_i>0$ for $\kappa=1$ (the green region in Fig.~\ref{fig:7d-all}).}
\begin{equation}\label{eq:7d-fq}
	\frac{p_1}{|g-1|} = - 2\sqrt2\frac{u\sqrt{(1-u)(2v-1)}}{\frac12+(2u-1)(2v-1)} 
	\, ,\qquad
	\frac{p_2}{|g-1|}= - 2\sqrt2\frac{v\sqrt{(1-v)(2u-1)}}{\frac12+(2u-1)(2v-1)} \,.
\end{equation} 
For $g\ge 2$, these conditions demand that two functions of $u$ and $v$ be rational. These solutions are then dense in the red region in Fig.~\ref{fig:7d-all}. For $g=0$, the case most interesting to us, $u$ and $v$ are discretised; all numerical solutions with $\sqrt{p_1^2+p_2^2}\le p_\mathrm{max}=50$ are shown in the green region of Fig.~\ref{fig:7d-all}. One of these dots is a solution found long ago \cite{Gauntlett:2002rv}.\footnote{In our language, these authors took the Ansatz $v=1$, $p_1=2$, $p_2=0$, justified by geometrical degenerations and similarity with the supersymmetric solutions; our equations reproduce their value $u=  (-3 + \sqrt{13})/4\simeq (6.6056)^{-1}$ in their (3.5). Their solution (3.11) has $\kappa=-1$ and is thus in the red region: in our language, it is $p_1=p_2=1$, $u^{-1}=v^{-1} \simeq 1.5536$.}
Upon increasing $p_\mathrm{max}$, solutions tend to accumulate more and more near the green/red boundary. This is simply because the denominators of \eqref{eq:7d-fq} vanish there.

A ``fan'' structure  emanates from the intersection of the green/red boundary and the diagonal gray line, as made clear by the zoomed-in plot in the right panel of Fig.~\ref{fig:7d-all}. (At the green/red boundary, $\kappa=0$.) Solving \eqref{eq:7d-fq} approximately we obtain 
\begin{equation}\label{eq:rays}
\begin{split}
	u \simeq \frac{3-\sqrt{3}}4 + \frac1{4p_1} \left((\sqrt3-1)(p_2-p_1)+1-2\sqrt3\right) \, ,\\
	v \simeq \frac{3+\sqrt{3}}4 + \frac1{4p_1} \left((\sqrt3+1)(p_2-p_1)-1-2\sqrt3\right)\,.
\end{split}
\end{equation}
Plotting these for fixed $t=p_2-p_1$ and increasing $p_1$ reproduces approximately the rays of the fan in the right panel of Fig.~\ref{fig:7d-all}. The $t=p_2-p_1=2$ ray lies on top of the BBBW gray diagonal line. This is the topological twist condition $c_1({\mathcal L}_1 \oplus {\mathcal L}_2)= 2g-2$ in \cite[(2.12)]{Bah:2012dg}, up to a sign in our definition of the $p_i$ (see the footnote above \eqref{eq:7d-fq}). This is needed for supersymmetry, but not for the equations of motion. Notice however that a similar computation shows that the total space $M_6$ of the $S^4$-fibration over $S^2$ is spin if and only if $t$ is even.

There are in fact similar, less clearly visible fans emanating from other points on the green/red boundary, which correspond to limits $p_1$, $p_2\to \infty$, $p_1/p_2 \in \mathbb{Q}$.  However, the original fan with $p_1=p_2\to \infty$ is special in that it lets one approach the extremal locus $u+v=3/2$. Recalling \eqref{eq:rho-7d}, near this locus we have $\rho-1 = \epsilon\ll 1$, which by \eqref{eq:Lambda4-probe} makes $\Lambda_4$ small. Notice that at the same time we are also obliged to approach the red/green boundary, which makes $R$ grow. We have checked that following one of the rays \eqref{eq:rays} we obtain $\Lambda_4\sim R^{-2}$, which is less favorable than \eqref{eq:Lambda-est}.

In any case, we are now going to see an even more severe problem with this class of solutions: they are perturbatively unstable.

\subsection{Perturbative instability} 
\label{sub:7d-inst}

\subparagraph{Maximal 7d supergravity.} While a full perturbative analysis would require a challenging KK reduction, just like for the flag solutions we can obtain part of the spectrum by a far more manageable dimensional reduction, this time performed from the 7d action in \cite{Liu:1999ai}. We find the potential
\begin{equation}\label{eq:V75}
	\begin{split}
	V=& \frac12m^2e^{-4\chi/3}(-8X_1X_2 -4X_1^{-1}X_2^{-2} -4X_1^{-2}X_2^{-1} +(X_1X_2)^{-4})\\ &-2\kappa e^{-10\chi/3} + 2e^{-16\chi/3} \left( q_1^2e^{-4\lambda_1} + q_2^2e^{-4\lambda_2}\right)
	\end{split}
\end{equation}
where $q_i= f_i R^2$, proportional to the flux quanta, are to be held fixed. Varying this potential with respect to the scalars $(\lambda_1, \lambda_2,\chi)$, using $V_0 = -12 L^{-2}_\mathrm{E}= -12 L^{-2}R^{-4/3}$, reproduces the equations of motion. (The subscript E denoting the 5d Einstein frame.) The kinetic term for these scalars is
\begin{equation}
	g= \left( \begin{array}{ccc}
		12 & 8 & 0 \\
		8 & 12 & 0 \\
		0 & 0 & 20/3
	\end{array}\right)\,.
\end{equation}
Multiplying its inverse by the Hessian of \eqref{eq:V75}, one obtains a matrix ${\mathcal M}$ related to the mass matrix by a similarity transformation, whose eigenvalues are then the physical masses. We obtain
\begin{equation}\label{eq:mass-max}
	L^2_\mathrm{E} {\mathcal M} = \frac8{5(2u+2v-1)} 
	\left(\begin{array}{ccc}
		-24uv+28u+28v-17 & -12(u-1)(2v-1) & -12(2u-1)(v-1) \\
		-4(2uv+u-4v+1) & -2(7u-2)(2v-1) & 4(3u-4)(v-1) \\
		-4(2uv-4u+v+1) & 4(u-1)(3v-4) & -2(2u-1)(7v-2) 
	\end{array}\right)\,.
\end{equation}
Unfortunately, this always has an eigenvalue $\le -4$, with equality only on the supersymmetric locus \eqref{eq:bbbw}.

\subparagraph{Minimal 7d supergravity.} This analysis can be repeated for the minimal 7d solutions we mentioned at the end of \ref{sub:7dsol}. However, we can also take advantage of the fact that this theory can be obtained from the maximal one upon setting  $X_1=X_2=X$, which implies $u=v=X^5$. Along this locus, one of the eigenvectors of \eqref{eq:mass-max} corresponds to the scalar direction $\lambda_1-\lambda_2$ and can be discarded, since this scalar is not included in this theory. The remaining two are
\begin{equation}
	L^2_\mathrm{E} m_\pm^2= \frac{4}{4X^5-1} \left( -(8X^{10}-10X^5+1) \pm (2X^5-1)\sqrt{16X^{10}-40X^5+49} \right)\,.
\end{equation}
Now there is no instability for $X^5 \le \frac{5+\sqrt{5}}8$. Thus a priori the uplifts of these solutions to massive IIA might be perturbatively stable. As we remarked earlier, they do suffer from the moduli problem, given that they only make sense for $\kappa=-1$, $g\ge 2$.



\section{Discussion}
In this paper, we established the dark bubble scenario realised by unstable M5-branes in M-theory compactifications. 

Our main results are non-supersymmetric solutions whose internal space is the flag manifold $F_6$. After determining the volume-minimising two-cycles in our flag geometry, we investigated the stability of M5-branes wrapping those cycles. We found that for the ``boomerang'' family there are regions in solution space where the M5-branes are stable, and others where they are not (see Fig.~\ref{fig:stab plane}).\footnote{Although we have identified unstable M-theory solutions, it would be interesting to see whether the non-supersymmetric solutions protected against M5-brane nucleation, i.e.~the uncoloured solutions in Fig.~\ref{fig:stab plane}, are truly stable and serve as counterexamples for the non-supersymmetric AdS conjecture \cite{Ooguri:2016pdq}. We leave this for the future.} It is important to note that the extremal point in the solution space, i.e.~where the M5-stability bound is saturated, does not correspond to an actual solution, as the flux quantisation conditions cannot be satisfied. 
In order to establish the dark bubble scenario, we rely on the Dirichlet approximation theorem by choosing a genuine but unstable flux vacuum sufficiently close to the extremal point. This allows for the realisation of the dark bubble scenario on the wrapped M5-brane bubble with an effective four-dimensional dark energy density satisfying $\rho_\Lambda\sim L^{-4} \sim R^{-4}\sim m_\mathrm{KK}^4$, where $L$ and $R$ are the AdS and internal radius respectively. We remark that a similar relation has been found in the dark bubble model of \cite{Danielsson:2022lsl,Danielsson:2023alz}; as an amusing parallel, it is also proposed in the dark dimension scenario, albeit the nature of the fifth extra dimension is different. 

As a consequence of this relation, this flag model can reproduce values of $\Lambda_4$ close to observations, with extra dimensions of the order of tens of microns, which lies in the range currently probed by sub-millimetre gravity experiments. Further work would be needed to establish if the model is really viable. In particular this would require the introduction of matter fields, and controlling their interactions to the KK modes.

We also considered some less successful alternatives. In the appendix we consider internal spaces of the form ${\rm KE}_4\times \Sigma_g$, finding that M5s are unstable only if $g\ge 2$, leading to moduli that should acquire a potential via higher-derivative corrections. We also looked at non-supersymmetric AdS$_5$ solutions of seven-dimensional maximal gauged supergravity \cite{Liu:1999ai} on a Riemann surface, which has a consistent uplift to M-theory \cite{Cvetic:1999xp}. We focussed primarily on a family of solutions where the Riemann surface is a two-sphere, $\Sigma=S^2$, and we found that the extremality bound for M5-branes wrapping it is saturated for precisely the supersymmetric solutions of \cite{Bah:2012dg}. On this supersymmetric locus, one of the scalar modes saturates the BF bound. However, we find that this mode violates the BF bound in all the non-supersymmetric solutions, thereby triggering perturbative instabilities. 
We have checked that such zero-mode instabilities are absent in the flag manifold construction, but a full  KK analysis would of course be needed to establish this rigorously. 

Keeping such caveats in mind, we conclude that the flag manifold construction realises the dark bubble scenario with a slightly super-extremal brane, relying only on tree-level ingredients and without the need of higher-curvature corrections as in refs.~\cite{Danielsson:2022lsl,Danielsson:2023alz}, which are notoriously hard to exhaustively control.\footnote{Regarding this issue, a more speculative interesting direction could be to consider a non-abelian realisation of the dark bubble scenario. Non-abelian bubbles in ten-dimensional curved space have been constructed in \cite{Menet:2025mbi}. Not only do the non-abelian $\alpha'$ corrections  bring an extremal abelian bubble to be slightly super-extremal, but these corrections are also parametrically larger than their abelian counterparts. This could therefore potentially provide some control over the various abelian $\alpha'$ corrections.}

\section*{Acknowledgments}
We thank U.~Danielsson, S.~Giri, A.~Karch for helpful discussions. VM and AT were supported by the INFN, and by the MUR-PRIN contract 2022YZ5BA2. VVH acknowledges financial support from Kungliga Fysiografiska sällskapet i Lund, and from the COST Action CA22113 through the STSM travel grant E-COST-GRANT-CA22113-37ae5938 to the University of Milano-Bicocca where this work was initiated.

\appendix

\section{K\"ahler products}
\label{sec:kahler}

 In this Appendix, we consider internal geometries that are products of K\"ahler spaces. In practice, the relevant internal spaces will always include at least one Riemann surface. As discussed in the main text, the moduli of these solutions might receive higher-derivative contributions to their potentials. Its minima could lie at degenerate points of the moduli space, where the Riemann surface is singular. For this reason, we only briefly present these solutions and the behaviour of their associated M5-branes.

We first present the case where the M5-branes are embedded in AdS$_5\times\text{K\"ahler}_4\times\Sigma$ solutions, with $\text{K\"ahler}_4$ a four-dimensional K\"ahler space, and $\Sigma$ a Riemann surface.

We start with the following Ansatz for the four-form flux
\begin{equation}
    G=\frac{g_1}{2}J_{\text{K}}\wedge J_{\text{K}}+g_2 J_{\text{K}}\wedge J_\Sigma\, ,
\end{equation}
with $J_{\text{K}}$ and $J_\Sigma$ the K\"ahler forms on $\text{K\"ahler}_4$  and the Riemann surface, respectively. This Ansatz ensures that the flux equations of motion $\dd G=0,\, \dd\ast G+\frac{1}{2}G\wedge G=0$ are satisfied, via the closure of the K\"ahler forms. 

The Einstein equations $R_{MN}=\frac{1}{2}\iota_MG\cdot\iota_NG-\frac{1}{6}g_{MN}|G|^2$ force the four-dimensional K\"ahler space to also be Einstein, so we introduce the parametrisation
\begin{equation}
  R_{\mu\nu}=-\frac{4}{L^2} g_{\mu\nu}, \qquad R_{mn}=\frac{\lambda_K}{L^2} g_{mn},\qquad  R_{ij}=\frac{\lambda_\Sigma}{L^2} g_{ij}, 
\end{equation}
with $L$ the AdS$_5$ length scale, and the above indices spanning AdS$_5$, the four-dimensional K\"ahler space, and the Riemann surface respectively.

The solutions to the equations of motion can be expressed in terms of the curvature of the four-dimensional K\"ahler space as 
\begin{equation}
    g_1=\pm \frac{2}{L}\sqrt{\lambda_K-2}\,,\qquad g_2=\pm\frac{\sqrt{2}}{L}\sqrt{8-\lambda_K}\,,\qquad\lambda_\Sigma=2(6-\lambda_K).
\end{equation}
We therefore have $\lambda_K\in[2,8]$, and the Riemann surface is a sphere when $\lambda_K\in[2,6[$, a torus when $\lambda_K=6$, or a higher genus ($g\ge 2$) surface when $\lambda_K\in]6,8]$.

\begin{figure}[!ht]
   \centering
\includegraphics[width=0.7\linewidth]{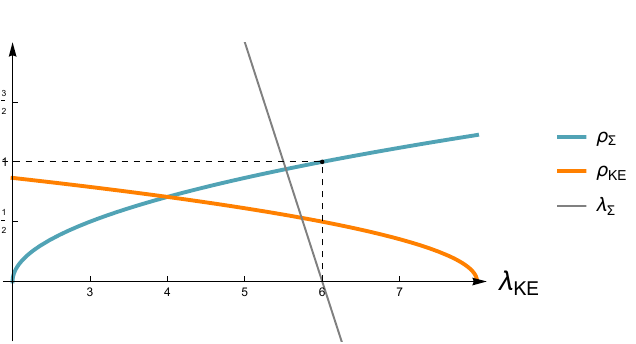} 
   \caption{\small The stability ratio for M5-branes wrapping calibrated cycles inside a K\"ahler$_4\times \Sigma$ geometry, in terms of the K\"ahler$_4$ curvature. The curvature of the Riemann surface is also displayed.}
    \label{Fig:KExSigma}
\end{figure}

We now consider M5-branes that are domain walls in AdS$_5$, wrapping the calibrated cycles $S_\Sigma$ or $S_K$:
\begin{equation}
    \text{Vol}(S_\Sigma)=\int_{S_\Sigma}J_\Sigma,\qquad\text{Vol}(S_K)=\int_{S_K}J_K.
\end{equation}
We calculate as before the stability ratio \eqref{eq:rho} for both types of M5-branes; they are displayed in Fig.~\ref{Fig:KExSigma}.
As we can see in Fig.~\ref{Fig:KExSigma}, an M5-brane wrapping a calibrated cycle in the K\"ahler$_4$ space is sub-extremal, so no M5-branes wrapping a two-cycle in K\"ahler$_4$ can expand.

On the other hand, an M5-brane wrapping the Riemann surface can be super-extremal. Unfortunately, this happens exactly when $\lambda_\Sigma < 0$ and thus $g\ge 2$, for which reason these models suffer from the moduli problem mentioned at the beginning of this appendix. 

The regime we care about for a dark bubble construction is that of M5-branes being slightly super-extremal, so for which the stability ratio is very slightly above one, and the deviation from extremality $\epsilon$ is very small. Again from Fig.~\ref{Fig:KExSigma}, we see that the smaller 
$\epsilon$ is, the smaller the curvature of the Riemann surface becomes. In other words, the smaller the cosmological constant, the larger the size of the extra dimensions. More precisely, we get $R\sim\Lambda^{-1/2}$ with $R=\frac{L}{\sqrt{|\lambda_\Sigma|}}$, recovering the same behaviour as the uplifts from seven dimensions in Sec.~\ref{sec:7d}.

Finally, we can consider M5-branes embedded in vacua with an internal space that is the product of three two-dimensional K\"ahler spaces.

We find that there are only two possible geometries: $S^2\times S^2\times S^2$ and $S^2\times S^2\times \Sigma$, with $\Sigma$ a two-torus or a higher genus Riemann surface. (This is in agreement with the results in \cite{Wulff:2016vqy}.)

We do not comment further on these possibilities since we find that only the  $S^2\times S^2\times \Sigma$ geometry allows for the nucleation of an M5-brane, with a behaviour exactly mirroring that of the $\text{K\"ahler}_4\times\Sigma$  solutions.

\bibliographystyle{JHEP}
\bibliography{biblio}

@article{Dolan:1984hv,
    author = "Dolan, B.",
    title = "{A New Solution of D = 11 Supergravity with Internal Isometry Group ${\rm SU}(3)\times {\rm SU}(2)\times {\rm U}(1)$}",
    doi = "10.1016/0370-2693(84)90758-5",
    journal = "Phys. Lett. B",
    volume = "140",
    pages = "304--306",
    year = "1984"
}

@article{Pope:1988xj,
    author = "Pope, C. N. and van Nieuwenhuizen, P.",
    title = "{Compactifications of $d=11$ Supergravity on K\"ahler Manifolds}",
    reportNumber = "CTP-TAMU-71/88, ITP-SB-88-63",
    doi = "10.1007/BF01257417",
    journal = "Commun. Math. Phys.",
    volume = "122",
    pages = "281",
    year = "1989"
}

@article{Martin:2008pf,
    author = "Martin, Jonathan E. and Reall, Harvey S.",
    title = "{On the stability and spectrum of non-supersymmetric $\mathrm{AdS}_5$ solutions of M-theory compactified on K\"ahler--Einstein spaces}",
    eprint = "0810.2707",
    archivePrefix = "arXiv",
    primaryClass = "hep-th",
    doi = "10.1088/1126-6708/2009/03/002",
    journal = "JHEP",
    volume = "03",
    pages = "002",
    year = "2009"
}

@article{Tomasiello:2007eq,
    author = "Tomasiello, Alessandro",
    title = "{New string vacua from twistor spaces}",
    eprint = "0712.1396",
    archivePrefix = "arXiv",
    primaryClass = "hep-th",
    doi = "10.1103/PhysRevD.78.046007",
    journal = "Phys. Rev. D",
    volume = "78",
    pages = "046007",
    year = "2008"
}

@article{Koerber:2008rx,
    author = "Koerber, Paul and Lust, Dieter and Tsimpis, Dimitrios",
    title = "{Type IIA AdS$_4$ compactifications on cosets, interpolations and domain walls}",
    eprint = "0804.0614",
    archivePrefix = "arXiv",
    primaryClass = "hep-th",
    reportNumber = "MPP-2008-29, LMU-ASC-19-08",
    doi = "10.1088/1126-6708/2008/07/017",
    journal = "JHEP",
    volume = "07",
    pages = "017",
    year = "2008"
}

@article{Menet:2025nbf,
    author = "Menet, Vincent and Tomasiello, Alessandro",
    title = "{Stability of non-supersymmetric vacua from calibrations}",
    eprint = "2507.02787",
    archivePrefix = "arXiv",
    primaryClass = "hep-th",
    doi = "10.1007/JHEP11(2025)070",
    journal = "JHEP",
    volume = "11",
    pages = "070",
    year = "2025"
}

@book{Tomasiello:2022dwe,
    author = "Tomasiello, Alessandro",
    title = "{Geometry of string theory compactifications}",
    doi = "10.1017/9781108635745",
    isbn = "978-1-108-63574-5, 978-1-108-47373-6",
    publisher = "Cambridge University Press",
    month = "1",
    year = "2022"
}

@article{Wulff:2016vqy,
    author = "Wulff, Linus",
    title = "{All symmetric space solutions of eleven-dimensional supergravity}",
    eprint = "1611.06139",
    archivePrefix = "arXiv",
    primaryClass = "hep-th",
    reportNumber = "IMPERIAL-TP-LW-2016-04",
    doi = "10.1088/1751-8121/aa70b6",
    journal = "J. Phys. A",
    volume = "50",
    number = "24",
    pages = "245401",
    year = "2017"
}

@article{Raucci:2025bev,
    author = "Raucci, Salvatore and Tomasiello, Alessandro",
    title = "{AdS} Vacua of Non-Supersymmetric Strings",
    eprint = "2510.01324",
    archivePrefix = "arXiv",
    primaryClass = "hep-th",
    reportNumber = "IFT-UAM/CSIC-25-101",
    doi = "10.1007/JHEP12(2025)057",
    journal = "JHEP",
    volume = "12",
    pages = "057",
    year = "2025"
}

@article{Altavilla_2022,
   title={Twistor geometry of the flag manifold},
   eprint = "2112.11100",
   volume={303},
   ISSN={1432-1823},
   url={http://dx.doi.org/10.1007/s00209-022-03161-x},
   DOI={10.1007/s00209-022-03161-x},
   number={1},
   journal={Mathematische Zeitschrift},
   publisher={Springer Science and Business Media LLC},
   author={Altavilla, Amedeo and Ballico, Edoardo and Brambilla, Maria Chiara and Salamon, Simon},
   year={2022},
   month=dec }

@misc{billey-gao-pawlowski,
      title={Introduction to the Cohomology of the Flag Variety}, 
      author={Sara C. Billey and Yibo Gao and Brendan Pawlowski},
      year={2025},
      eprint={2506.21064},
      archivePrefix={arXiv},
      primaryClass={math.CO},
      url={https://arxiv.org/abs/2506.21064}, 
}

@article{Boya:2003km,
    author = "Boya, L. J. and Sudarshan, G. and Tilma, Todd Edward",
    title = "{Volumes of compact manifolds}",
    eprint = "math-ph/0210033",
    archivePrefix = "arXiv",
    doi = "10.1016/S0034-4877(03)80038-1",
    journal = "Rept. Math. Phys.",
    volume = "52",
    pages = "401--422",
    year = "2003"
}

@article{Ooguri:2017njy,
    author = "Ooguri, Hirosi and Spodyneiko, Lev",
    title = "{New Kaluza--Klein instantons and the decay of AdS vacua}",
    eprint = "1703.03105",
    archivePrefix = "arXiv",
    primaryClass = "hep-th",
    reportNumber = "CALT-TH-2017-014, IPMU-17-0041",
    doi = "10.1103/PhysRevD.96.026016",
    journal = "Phys. Rev. D",
    volume = "96",
    number = "2",
    pages = "026016",
    year = "2017"
}

@article{Coleman:1980aw,
    author = "Coleman, Sidney R. and De Luccia, Frank",
    title = "Gravitational effects on and of vacuum Decay",
    reportNumber = "SLAC-PUB-2463",
    doi = "10.1103/PhysRevD.21.3305",
    journal = "Phys. Rev. D",
    volume = "21",
    pages = "3305",
    year = "1980"
}

@article{Brown:1988kg,
    author = "Brown, J. David and Teitelboim, C.",
    title = "Neutralization of the Cosmological Constant by Membrane Creation",
    doi = "10.1016/0550-3213(88)90559-7",
    journal = "Nucl. Phys. B",
    volume = "297",
    pages = "787--836",
    year = "1988"
}

@article{Lu:1999bc,
      author         = "Lu, Hong and Pope, C.N.",
      title          = "{Exact embedding of ${\mathcal{N}}\!=1$, $D = 7$ gauged supergravity in $D = 11$}",
      journal        = "Phys.Lett.",
      volume         = "B467",
      pages          = "67-72",
      doi            = "10.1016/S0370-2693(99)01170-3",
      year           = "1999",
      eprint         = "hep-th/9906168",
      archivePrefix  = "arXiv",
      primaryClass   = "hep-th",
      reportNumber   = "CTP-TAMU-26-99, UPR-850-T, SISSA-REF-68-99-EP",
      SLACcitation   = "%%CITATION = HEP-TH/9906168;%%",
}

@article{Passias:2015gya,
    author = "Passias, Achilleas and Rota, Andrea and Tomasiello, Alessandro",
    title = "Universal Consistent Truncation for 6D/7D Gauge/Gravity Duals",
    eprint = "1506.05462",
    archivePrefix = "arXiv",
    primaryClass = "hep-th",
    doi = "10.1007/JHEP10(2015)187",
    journal = "JHEP",
    volume = "10",
    pages = "187",
    year = "2015"
}

@article{Townsend:1983kk,
      author         = "Townsend, P.K. and van Nieuwenhuizen, P.",
      title          = "Gauged Seven-Dimensional Supergravity",
      journal        = "Phys.Lett.",
      volume         = "B125",
      pages          = "41-46",
      doi            = "10.1016/0370-2693(83)91230-3",
      year           = "1983",
      reportNumber   = "CERN-TH-3508",
      SLACcitation   = "%%CITATION = PHLTA,B125,41;%%",
}

@article{Pernici:1984xx,
    author = "Pernici, M. and Pilch, K. and van Nieuwenhuizen, P.",
    title = "Gauged Maximally Extended Supergravity in Seven Dimensions",
    reportNumber = "ITP-SB-84-31",
    doi = "10.1016/0370-2693(84)90813-X",
    journal = "Phys. Lett. B",
    volume = "143",
    pages = "103--107",
    year = "1984"
}

@article{Cvetic:1999xp,
    author = "Cveti{\v c}, Mirjam and Duff, M. J. and Hoxha, P. and Liu, James T. and Lu, Hong and Lu, J. X. and Mart{\'\i ne}z-Acosta, R. and Pope, C. N. and Sati, H. and Tran, Tuan A.",
    title = "Embedding {AdS} Black Holes in Ten Dimensions and Eleven Dimensions",
    eprint = "hep-th/9903214",
    archivePrefix = "arXiv",
    reportNumber = "UPR-0840-T, CTP-TAMU-11-99, RU-99-4-B",
    doi = "10.1016/S0550-3213(99)00419-8",
    journal = "Nucl. Phys. B",
    volume = "558",
    pages = "96--126",
    year = "1999"
}

@article{Liu:1999ai,
    author = "Liu, James T. and Minasian, Ruben",
    title = "Black Holes and Membranes in Ad$S_7$",
    eprint = "hep-th/9903269",
    archivePrefix = "arXiv",
    reportNumber = "RU-99-5-B, YCTP-P9-99",
    doi = "10.1016/S0370-2693(99)00500-6",
    journal = "Phys. Lett. B",
    volume = "457",
    pages = "39--46",
    year = "1999"
}

@article{Maldacena:2000mw,
    author = "Maldacena, Juan Martin and Nunez, Carlos",
    editor = "Duff, Michael J. and Liu, J. T. and Lu, J.",
    title = "Supergravity Description of Field Theories on Curved Manifolds and a No Go Theorem",
    eprint = "hep-th/0007018",
    archivePrefix = "arXiv",
    doi = "10.1142/S0217751X01003937",
    journal = "Int. J. Mod. Phys. A",
    volume = "16",
    pages = "822--855",
    year = "2001"
}

@article{Bah:2012dg,
    author = "Bah, Ibrahima and Beem, Christopher and Bobev, Nikolay and Wecht, Brian",
    title = "{Four-dimensional SCFTs from M5-branes}",
    eprint = "1203.0303",
    archivePrefix = "arXiv",
    primaryClass = "hep-th",
    reportNumber = "MCTP-12-05, QMUL-PH-12-05",
    doi = "10.1007/JHEP06(2012)005",
    journal = "JHEP",
    volume = "06",
    pages = "005",
    year = "2012"
}

@article{LeKhongVan,
  author   = {L{\^{e}} H{\`o}ng V{\^{a}}n},
  title    = {Minimal surfaces in homogeneous spaces},
  journal  = {Mathematics of the USSR-Izvestiya},
  year     = {1989},
  volume   = {32},
  number   = {2},
  pages    = {413--436},
  doi      = {10.1070/IM1989v032n02ABEH000773},
  url      = {https://doi.org}
}

@article{Basile:2025lek,
    author = "Basile, Ivano and Borys, Alessandro and Masias, Joaquin",
    title = "{Dynamical dark energy in 0{\textquoteright}B braneworlds}",
    eprint = "2502.20438",
    archivePrefix = "arXiv",
    primaryClass = "hep-th",
    doi = "10.1140/epjc/s10052-025-14706-9",
    journal = "Eur. Phys. J. C",
    volume = "85",
    number = "9",
    pages = "986",
    year = "2025"
}

@article{Danielsson:2022lsl,
    author = "Danielsson, Ulf and Henriksson, Oscar and Panizo, Daniel",
    title = "{Stringy realization of a small and positive cosmological constant in dark bubble cosmology}",
    eprint = "2211.10191",
    archivePrefix = "arXiv",
    primaryClass = "hep-th",
    reportNumber = "UUITP - 49/22, HIP-2022-29/TH",
    doi = "10.1103/PhysRevD.107.026020",
    journal = "Phys. Rev. D",
    volume = "107",
    number = "2",
    pages = "026020",
    year = "2023"
}

@article{Danielsson:2023alz,
    author = "Danielsson, Ulf and Panizo, Daniel",
    title = "{Experimental tests of dark bubble cosmology}",
    eprint = "2311.14589",
    archivePrefix = "arXiv",
    primaryClass = "hep-th",
    doi = "10.1103/PhysRevD.109.026003",
    journal = "Phys. Rev. D",
    volume = "109",
    number = "2",
    pages = "026003",
    year = "2024"
}

@article{Basile:2025lwx,
    author = "Basile, Ivano and Borys, Alessandro and Masias, Joaquin",
    title = "{Dark bubble cosmology and the equivalence principle}",
    eprint = "2507.03748",
    archivePrefix = "arXiv",
    primaryClass = "hep-ph",
    doi = "10.1103/qm3b-kpm8",
    journal = "Phys. Rev. D",
    volume = "113",
    number = "2",
    pages = "026009",
    year = "2026"
}

@article{Danielsson:2024frw,
    author = "Danielsson, Ulf and Van Hemelryck, Vincent",
    title = "{Charged Nariai black holes on the dark bubble}",
    eprint = "2405.13679",
    archivePrefix = "arXiv",
    primaryClass = "hep-th",
    reportNumber = "UUITP-13/24",
    doi = "10.1088/1361-6382/ad8f8d",
    journal = "Class. Quant. Grav.",
    volume = "41",
    number = "24",
    pages = "245011",
    year = "2024"
}

@article{Danielsson:2026qch,
    author = "Danielsson, Ulf and Panizo, Daniel and Van Hemelryck, Vincent",
    title = "{Self-gravitating electromagnetic waves in the dark bubble model}",
    eprint = "2606.16547",
    archivePrefix = "arXiv",
    primaryClass = "hep-th",
    month = "6",
    year = "2026"
}

@article{Basile:2023tvh,
    author = "Basile, Ivano and Danielsson, Ulf and Giri, Suvendu and Panizo, Daniel",
    title = "{Shedding light on dark bubble cosmology}",
    eprint = "2310.15032",
    archivePrefix = "arXiv",
    primaryClass = "hep-th",
    reportNumber = "UUITP-30/23",
    doi = "10.1007/JHEP02(2024)112",
    journal = "JHEP",
    volume = "02",
    pages = "112",
    year = "2024"
}

@article{Banerjee:2018qey,
    author = "Banerjee, Souvik and Danielsson, Ulf and Dibitetto, Giuseppe and Giri, Suvendu and Schillo, Marjorie",
    title = "{Emergent de Sitter cosmology from decaying Anti{\textendash}de Sitter space}",
    eprint = "1807.01570",
    archivePrefix = "arXiv",
    primaryClass = "hep-th",
    reportNumber = "UUITP-27-18",
    doi = "10.1103/PhysRevLett.121.261301",
    journal = "Phys. Rev. Lett.",
    volume = "121",
    number = "26",
    pages = "261301",
    year = "2018"
}

@article{Romans:1984an,
    author = "Romans, L. J.",
    title = "{New Compactifications of Chiral $N=2 d=10$ Supergravity}",
    reportNumber = "CALT-68-1204",
    doi = "10.1016/0370-2693(85)90479-4",
    journal = "Phys. Lett. B",
    volume = "153",
    pages = "392--396",
    year = "1985"
}

@article{Danielsson:2026wqp,
    author = "Danielsson, Ulf and Giri, Suvendu",
    title = "{Dark bubbles, dark dimensions and fat gravitons}",
    eprint = "2606.20942",
    archivePrefix = "arXiv",
    primaryClass = "hep-th",
    reportNumber = "UUITP-14/26",
    month = "6",
    year = "2026"
}

@article{Banerjee:2022ree,
    author = "Banerjee, Souvik and Danielsson, Ulf and Giri, Suvendu",
    title = "{Features of a dark energy model in string theory}",
    eprint = "2212.14004",
    archivePrefix = "arXiv",
    primaryClass = "hep-th",
    reportNumber = "UUITP-63/22",
    doi = "10.1103/PhysRevD.108.126009",
    journal = "Phys. Rev. D",
    volume = "108",
    number = "12",
    pages = "126009",
    year = "2023"
}

@article{Danielsson:2025ffk,
    author = "Danielsson, Ulf and Giri, Suvendu",
    title = "{Weak gravity at micron scales from dark bubble cosmology and its cosmological consequences}",
    eprint = "2511.21362",
    archivePrefix = "arXiv",
    primaryClass = "hep-th",
    reportNumber = "UUITP-36/25",
    doi = "10.1103/819l-m95m",
    journal = "Phys. Rev. D",
    volume = "113",
    number = "12",
    pages = "126010",
    year = "2026"
}

@article{Banerjee:2023uto,
    author = "Banerjee, Souvik and Danielsson, Ulf and Zemsch, Maximilian",
    title = "{The dark bubbleography}",
    eprint = "2311.16242",
    archivePrefix = "arXiv",
    primaryClass = "hep-th",
    doi = "10.1007/JHEP02(2024)102",
    journal = "JHEP",
    volume = "02",
    pages = "102",
    year = "2024"
}

@article{Banerjee:2019fzz,
    author = "Banerjee, Souvik and Danielsson, Ulf and Dibitetto, Giuseppe and Giri, Suvendu and Schillo, Marjorie",
    title = "{de Sitter cosmology on an expanding bubble}",
    eprint = "1907.04268",
    archivePrefix = "arXiv",
    primaryClass = "hep-th",
    reportNumber = "UUITP-26/19",
    doi = "10.1007/JHEP10(2019)164",
    journal = "JHEP",
    volume = "10",
    pages = "164",
    year = "2019"
}

@article{Banerjee:2020wix,
    author = "Banerjee, Souvik and Danielsson, Ulf and Giri, Suvendu",
    title = "{Dark bubbles: decorating the wall}",
    eprint = "2001.07433",
    archivePrefix = "arXiv",
    primaryClass = "hep-th",
    reportNumber = "UUITP-1/20",
    doi = "10.1007/JHEP04(2020)085",
    journal = "JHEP",
    volume = "04",
    pages = "085",
    year = "2020"
}

@article{Banerjee:2020wov,
    author = "Banerjee, Souvik and Danielsson, Ulf and Giri, Suvendu",
    title = "{Bubble needs strings}",
    eprint = "2009.01597",
    archivePrefix = "arXiv",
    primaryClass = "hep-th",
    reportNumber = "UUITP-33/20",
    doi = "10.1007/JHEP03(2021)250",
    journal = "JHEP",
    volume = "21",
    pages = "250",
    year = "2020"
}

@article{Montero:2022prj,
    author = "Montero, Miguel and Vafa, Cumrun and Valenzuela, Irene",
    title = "{The dark dimension and the Swampland}",
    eprint = "2205.12293",
    archivePrefix = "arXiv",
    primaryClass = "hep-th",
    doi = "10.1007/JHEP02(2023)022",
    journal = "JHEP",
    volume = "02",
    pages = "022",
    year = "2023"
}

@article{Randall:1999vf,
    author = "Randall, Lisa and Sundrum, Raman",
    title = "{An alternative to compactification}",
    eprint = "hep-th/9906064",
    archivePrefix = "arXiv",
    reportNumber = "MIT-CTP-2874, PUPT-1867, BUHEP-99-13",
    doi = "10.1103/PhysRevLett.83.4690",
    journal = "Phys. Rev. Lett.",
    volume = "83",
    pages = "4690--4693",
    year = "1999"
}

@article{Karch:2000ct,
    author = "Karch, Andreas and Randall, Lisa",
    editor = "Duff, Michael J. and Liu, J. T. and Lu, J.",
    title = "{Locally localized gravity}",
    eprint = "hep-th/0011156",
    archivePrefix = "arXiv",
    reportNumber = "MIT-CTP-3099",
    doi = "10.1088/1126-6708/2001/05/008",
    journal = "JHEP",
    volume = "05",
    pages = "008",
    year = "2001"
}

@article{Ooguri:2016pdq,
    author = "Ooguri, Hirosi and Vafa, Cumrun",
    title = "{Non-supersymmetric AdS and the Swampland}",
    eprint = "1610.01533",
    archivePrefix = "arXiv",
    primaryClass = "hep-th",
    reportNumber = "CALT-TH-2016-027, IPMU16-0139",
    doi = "10.4310/ATMP.2017.v21.n7.a8",
    journal = "Adv. Theor. Math. Phys.",
    volume = "21",
    pages = "1787--1801",
    year = "2017"
}

@article{Imaanpur:2012jx,
    author = "Imaanpur, Ali",
    title = "A New Solution of Eleven-dimensional Supergravity",
    eprint = "1205.1349",
    archivePrefix = "arXiv",
    primaryClass = "hep-th",
    doi = "10.1088/0264-9381/30/6/065021",
    journal = "Class. Quant. Grav.",
    volume = "30",
    pages = "065021",
    year = "2013"
}

@article{Gauntlett:2002rv,
    author = "Gauntlett, Jerome P. and Kim, Nakwoo and Pakis, Stathis and Waldram, Daniel",
    title = "{M-theory solutions with AdS factors}",
    eprint = "hep-th/0202184",
    archivePrefix = "arXiv",
    reportNumber = "QMUL-PH-02-04",
    doi = "10.1088/0264-9381/19/15/305",
    journal = "Class. Quant. Grav.",
    volume = "19",
    pages = "3927--3946",
    year = "2002"
}

@article{Gaiotto:2009gz,
    author = "Gaiotto, Davide and Maldacena, Juan",
    title = "The Gravity Duals of $\mathcal{N}=2$ Superconformal Field Theories",
    eprint = "0904.4466",
    archivePrefix = "arXiv",
    primaryClass = "hep-th",
    doi = "10.1007/JHEP10(2012)189",
    journal = "JHEP",
    volume = "10",
    pages = "189",
    year = "2012"
}

@article{Danielsson:2022fhd,
	archiveprefix = {arXiv},
	author = {Danielsson, Ulf and Panizo, Daniel and Tielemans, Rob},
	doi = {10.1103/PhysRevD.106.024002},
	eprint = {2202.00545},
	journal = {Phys. Rev. D},
	number = {2},
	pages = {024002},
	primaryclass = {hep-th},
	reportnumber = {UUITP - 03/22},
	title = {{Gravitational waves in dark bubble cosmology}},
	volume = {106},
	year = {2022}}

@article{Banerjee:2021qei,
	archiveprefix = {arXiv},
	author = {Banerjee, Souvik and Danielsson, Ulf and Giri, Suvendu},
	doi = {10.1007/JHEP09(2021)158},
	eprint = {2102.02164},
	journal = {JHEP},
	pages = {158},
	primaryclass = {hep-th},
	reportnumber = {UUITP-05/21},
	title = {{Dark bubbles and black holes}},
	volume = {09},
	year = {2021}}

@article{Menet:2025mbi,
    author = "Menet, Vincent and Tomasiello, Alessandro",
    title = "{A weakly non-abelian decay channel}",
    eprint = "2512.09884",
    archivePrefix = "arXiv",
    primaryClass = "hep-th",
    doi = "10.1007/JHEP07(2026)142",
    journal = "JHEP",
    volume = "07",
    pages = "142",
    year = "2026"
}

\end{document}